\documentclass[aps,twocolumn,prb,10pt,showpacs,showkeys]{revtex4-1}

\usepackage{amsmath,amssymb}
\usepackage{graphicx}
\usepackage{dcolumn}
\usepackage{bm}
\usepackage{multirow}
\usepackage{float}
\usepackage{subfigure}
\usepackage{color}
\usepackage{ulem}

\begin{document}

\title{Electron transport through a single-molecule junction with multiple pathways under time-periodic fields: A Floquet-scattering formalism}

\author{Liang-Yan Hsu}
\email{lianghsu@princeton.edu}
\author{Herschel Rabitz}
\email{hrabitz@princeton.edu}
\affiliation{Department of Chemistry, 
Princeton University, Princeton, New Jersey 08544, USA}


\begin{abstract} 
We develop a new general formulation to explore light-driven electron transport through a single-molecule device with multiple pathways. Three individual systems are proposed including (i) a two-terminal molecular junction based on phenyl-acetylene macrocycle (PAM), (ii) PAM with three terminals, and (iii) a parallel molecular circuit. The computations show that PAM-based optoelectronic switches have robust large on-off ratios and weak-field operating conditions, which are not sensitive to asymmetric molecule-lead couplings. In addition, field-amplitude power laws for one- and two-photon assisted tunneling are evident in the computational results, and the laws can be proven by using perturbation analysis. For PAM-based optoelectronic routers, we show that it is possible to manipulate the direction of electric current through the PAM molecule by using a weak linearly polarized laser field. For parallel circuits made of molecular quantum dots, the condition of coherent destruction of tunneling is derived by using the rotating wave approximation and the high-frequency approximation.   
\end{abstract}

\maketitle

\section{Introduction}
Molecular electronics is an active field with many potential applications and novel innovations in nanoelectronic devices \cite{molecular_electronics1,molecular_electronics2,molecular_electronics3,molecular_electronics4,molecular_electronics4_1,molecular_electronics5,molecular_electronics6}. An ultimate goal in molecular electronics is to construct molecule-based integrated circuits. To achieve this goal, understanding electron transport through a single-molecule device (junction) is a required first step. During the past two decades, great experimental progress has been made on single-molecule junctions due to advancements in microfabrication and self-assembly techniques \cite{Science.278.252,Nature.407.57}. In addition, improvements in theoretical modeling have enabled successful explanations and predictions of experimental observations, e.g., Kondo effect\cite{Kondo1,Kondo2}, thermoelectricity \cite{thermal}, orbital gating \cite{gating}, destructive quantum interference \cite{cross2,cross3,DQI_benzene}, and conductance dependence upon molecular conformation \cite{conformation1,conformation2,conformation3}.  The Landauer formula combined with the non-equilibrium Green's function method \cite{quantum_transport1, quantum_transport2,NEGFDFT0,NEGFDFT1,NEGFDFT2,NEGFDFT3} has become a common approach to analyze electron transport through a single-molecule junction in the absence of applied fields. For a practical device, the capability of manipulating current by applying such fields is necessary. Therefore, control of electron transport, e.g., by means of light \cite{opticalgate} or gate electrodes \cite{Kondo1,gating}, is a prime issue for molecular electronics.

 
Light is a potential control tool for operating ultrafast electronic devices due to a wide variety of control options, e.g., field strength, phase, frequency, and polarization, compared with a static field provided by gate electrodes. Since 1960s \cite{PAT1} electron transport in the presence of a coherent light field has attracted general interest due to photon assisted tunneling observed in mesoscopic systems exposed to microwave radiation \cite{Tien,multiphoton}. In superconductor-insulator-superconductor tunnel junctions \cite{CDT1,PAT1} and semiconductor nanostructures \cite{QD1,nanostructure1}, light-driven electron transport has been extensively studied for some time.

Recently, light-driven electron transport at the molecular level became an emerging field due to experimental improvements in molecular junctions \cite{exprectifier1}. Moreover, a wealth of coherent light sources in the visible and the infrared regime, compatible with electronic energy scales of molecular systems, motivated innovative theoretical studies in this field \cite{Transport_Floquet1,Transport_Floquet2,ratchet,rectifier,shot_noise_1,shot_noise_2,PAT1_Mujica,PAT2_Mujica,epcoupling,memory,optimalcontrol}. Within a time-dependent perturbation theory along with the extended H\"uckel model, photon-assisted tunneling was predicted in the presence of high-intensity fields \cite{Transport_Floquet1,Transport_Floquet2}. Theoretical studies using Floquet theory together with a tight-binding model have shown that molecular wires can act as coherent quantum ratchets \cite{ratchet}, and that current rectification \cite{rectifier}, shot noise control \cite{shot_noise_1,shot_noise_2}, photon-assisted tunneling \cite{PAT1_Mujica,PAT2_Mujica}, and coherent destruction of tunneling could be observed in a linear molecular wire. Vibrational effects \cite{epcoupling}, memory effects \cite{memory}, and optimal control of shot noise \cite{optimalcontrol} have been investigated in a two-level model. However, most studies focus on linear molecular wires or few-level models and neglect the nature of molecular structure, e.g., cross-conjugated molecules and polycyclic aromatic hydrocarbons, which offer multiple pathways for tunneling electrons.

Multiple pathways in a molecular junction can be regarded as forming a ``network'', which is of importance in molecular electronics for several reasons. First, such networks can be thought of as analogous to electric circuits, e.g., a double-backbone molecule corresponding to a parallel circuit. Based on Kirchhoff's circuit laws, the conductance of two parallel constituents in a circuit equals the sum of the conductance of the individuals. However, at the molecular scale Kirchhoff's circuit laws do not hold since the phase coherence length of tunneling electrons is comparable to the circuit scale, i.e., quantum interference effects play an important role. A conductance superposition law in a parallel circuit has been proposed \cite{parallel1,parallel2} and experimentally reported \cite{parallelexp}. Second, pathway differences in a network can lead to destructive quantum interference \cite{Interference1,Interference2,Interference3,Interference4,Interference5,parameter1,parameter2,benzene,Interference6,optoelec,SMERD}, which can be exploited to form single-molecule optoelectronic switches \cite{optoelec}, single-molecule electric revolving doors \cite{SMERD}, and quantum interference effect transistors \cite{Interference2,Interference3,parameter2}. Destructive quantum interference with tunneling has been experimentally observed in aromatic hydrocarbons \cite{DQI_benzene,ring} and cross-conjugated molecules \cite{cross2,cross3} at room temperature, indicating that molecular structure can be a key resource for coherent quantum transport. Third, a network such as a single molecule connected to multi-terminals \cite{multi1,multi2} could form a nanoscale router. In order to manipulate current in a molecular router, an understanding of the correlation between the molecular structure and current direction is required.

We present a new general formulation to explore laser-driven transport through three individual networks consisting of (i) a two-terminal molecular junction based on phenyl-acetylene macrocycle (PAM) \cite{PAM1,PAM2}, (ii) PAM with three terminals, and (iii) a parallel molecular circuit. The formulation generalizes a scattering analysis  based on single-particle Green’s functions within the framework of the non-Hermitian Floquet theory \cite{shot_noise_1,shot_noise_2}. The new formulation is valid for arbitrary strength as well as frequency of a driving field and enables the modeling of light-driven transport through multi-terminal and multi-orbital systems in terms of transmission probabilities. We will discuss field-amplitude power laws in the weak-field regime, the effects of asymmetric molecule-electrode couplings, and the influence of laser polarization on electron transport through PAM.



\section{Formulation}\label{section2}

\subsection{Model Hamiltonian}

Electron transport through a single-molecule junction in the presence of time-periodic fields can be described by the time-dependent Hamiltonian $H_{\textrm{tot}}(t)$ composed of the external-field-driven molecular Hamiltonian $H_{\textrm{mol}}(t)$, the lead (electrode) Hamiltonian $H_{\textrm{lead}}$, and the molecule-lead coupling term $H_{\textrm{coup}}$, i.e., $H_{\textrm{tot}}(t)=H_{\textrm{mol}}(t)+H_{\textrm{lead}}+H_{\textrm{coup}}$. To focus on the effect of molecular networks, we remove other possible effects, e.g., many-body interactions, and adopt a single-electron tight-binding model to describe $H_{\textrm{mol}}(t)$ 
\begin{align}
\label{Model_Hamiltonian1}
H_{\textrm{mol}}(t)=\sum_{nn'}H_{nn'}(t)a^\dagger_{n} a_{n'},
\end{align}
where $a_n$ ($a^{\dagger}_n$) are Fermion operators which annihilate (create) an electron in the atomic orbital $|n\rangle$ in the molecule. The field-driven molecular Hamiltonian satisfies $H_{\textrm{mol}}(t)=H_{\textrm{mol}}(t+T)$ due to the time-periodic field with frequency $\omega=2\pi/T$. Note that the field is not limited to lasers, and it can be any type of time-periodic resource. In Eq.~(\ref{Model_Hamiltonian1}) the atomic orbitals $n$ and $n'$ need not be nearest neighbors, i.e., long-range interactions between the atomic orbitals $n$ and $n'$ can be considered.

We use a noninteracting electron gas model to describe the leads 
\begin{align}
\label{Model_Hamiltonian2}
H_{\textrm{lead}}=\sum_{\alpha}^{\kappa} \sum_{q}\epsilon_{\alpha q} b^{\dagger}_{\alpha q} b_{\alpha q}, 
\end{align}	
where $\kappa$ is the total number of the electrodes and $b_{\alpha q}$ ($b^{\dagger}_{\alpha q}$) are Fermion operators which annihilate (create) an electron in the state $|\alpha q\rangle$ with energy $\epsilon_{\alpha q}$ in the lead $\alpha$. The molecule-lead coupling term is
\begin{align}
\label{Model_Hamiltonian3}
H_{\textrm{coup}}=\sum_{\alpha}^{\kappa} \sum_{q} \sum_{n} V_{\alpha q,n} b^{\dagger}_{\alpha q}a_{n} +H.c.,
\end{align}	
where $V_{\alpha q,n}$ denotes the electron transfer integrals between the atomic orbital $|n\rangle$ of the molecule and the orbital $|\alpha q\rangle$ in the lead. The coupling function in the energy domain is defined as 
\begin{align}
\label{self1}
\Gamma_{\alpha,n n'}(\epsilon)=2\pi\sum_{q}V^{*}_{\alpha q,n}V_{\alpha q,n'}\delta(\epsilon-\epsilon_{\alpha q}),
\end{align}
and its Fourier transform in the time domain is
\begin{align}
\label{self2}
\Gamma_{\alpha,n n'}(t)&=\frac{1}{\hbar}\frac{1}{2\pi}\int^{\infty}_{-\infty}d\epsilon\Gamma_{\alpha,n n'}(\epsilon)\times e^{-\frac{i\epsilon t}{\hbar}} \nonumber \\
&=\frac{1}{\hbar}\sum_{q}V^{*}_{\alpha q,n}V_{\alpha q,n'}e^{-\frac{i\epsilon_{\alpha q} t}{\hbar}}.
\end{align}
The off-diagonal terms of the coupling function originate from the interactions between $|\alpha q\rangle$ in the lead $\alpha$ and two different atomic orbitals on the molecule.
As $|\alpha q\rangle$ only interacts with a specific atomic orbital $|v\rangle$ on the molecule, i.e., $n=n'=v$, we can derive $\Gamma_{\alpha,vv}(\epsilon)=2\pi\sum_{q}|V_{\alpha q,v}|^2\delta(\epsilon-\epsilon_{\alpha q})$, consistent with the previous result \cite{shot_noise_1,optoelec}.  


\subsection{Fundamental Solution}

In the Heisenberg picture, the time-evolution operator $U(t,t_0)$ associated with the total system Hamiltonian $H_{\textrm{tot}}(t)$ is governed by the equation
\begin{align}
\label{evolution2_1}
i\hbar \frac{d U(t,t_0)}{d t}=H_{\textrm{tot}}(t)U(t,t_0),
\end{align} 
and the equations of motion of the electron annihilation operators can be expressed in terms of $H_{\textrm{tot}}(t)$ and $U(t,t_0)$ as follows
\begin{align}
\label{Heisenberg3_1}
i\hbar\frac{d a_n(t)}{dt}=-U^{\dagger}(t,t_0)[H_{\textrm{tot}}(t),a_n] U(t,t_0), \\
\label{Heisenberg3_2}
i\hbar\frac{d b_{\alpha q}(t)}{dt}=-U^{\dagger}(t,t_0)[H_{\textrm{tot}}(t),b_{\alpha q}]U(t,t_0).
\end{align}

By substitution of Eqs.~(\ref{Model_Hamiltonian1}) -- (\ref{Model_Hamiltonian3}) into Eqs.~(\ref{Heisenberg3_1}) and (\ref{Heisenberg3_2}), we obtain equations of motion for electron annihilation operators in a single-molecule junction
\begin{align}
\label{Heisenberg4}
\frac{d a_n(t)}{dt}&=-\frac{i}{\hbar}\sum_{n'}H_{nn'}(t)a_{n'}(t)-\frac{i}{\hbar}\sum_{\alpha q}V^{*}_{\alpha q,n}(t)b_{\alpha q}(t),  \nonumber \\
\frac{d b_{\alpha q}(t)}{dt}&=-\frac{i}{\hbar}\epsilon_{\alpha q}b_{\alpha q}(t)-\frac{i}{\hbar}\sum_{n}V_{\alpha q,n}(t)a_{n}(t).
\end{align}

The linear differential equations in Eq. (\ref{Heisenberg4}) have the formal solution \cite{shot_noise_1,shot_noise_2}
\begin{align}
\label{fundamental}
b_{\alpha q}(t)&=\sum_{n}\langle\alpha q|U(t,t_0)|n\rangle a_n(t_0) \nonumber \\
               &+\sum_{\alpha'}^{\kappa}\sum_{q'}\langle\alpha q|U(t,t_0)|\alpha' q'\rangle b_{\alpha' q'}(t_0),
\end{align}
where $b_{\alpha q}(t)$ is determined by the initial conditions $a_n(t_0)$ and $b_{\alpha' q'}(t_0)$. The electrodes are initially in a thermal equilibrium state described by the initial density matrix $\rho_0 = e^{-(H_{\textrm{leads}}-\sum_{\alpha}\mu_{\alpha}N_{\alpha})/k_{\textrm{B}}\theta}$, where $\theta$ is the temperature, $N_{\alpha}=\sum_{q}b^{\dagger}_{\alpha q}b_{\alpha q}$ is an electron number operator, and $\mu_{\alpha}$ is the chemical potential associated with the lead $\alpha$. As a result, the average electron occupation number at $t_0$ can be expressed as $\langle b^{\dagger}_{\alpha'q'}(t_0)b_{\alpha q}(t_0)\rangle=\delta_{\alpha'\alpha}\delta_{q'q}f_{\alpha}(\epsilon_{\alpha q})$, where $f_{\alpha}(\epsilon)=(1+e^{(\epsilon-\mu_{\alpha})/\textrm{k}_{\textrm{B}}\theta})^{-1}$ is the Fermi function of the lead $\alpha$. For a two terminal system with symmetric chemical potentials, the left and right chemical potentials can be respectively written as $\mu_{\textrm{L}}=\mu-\textrm{e}V_{\textrm{SD}}/2$ and $\mu_{\textrm{R}}=\mu+\textrm{e}V_{\textrm{SD}}/2$, in terms of the zero-bias chemical potential $\mu$, electric charge e, and the source-drain voltage $V_{\textrm{SD}}$.

\subsection{Electric Current and Rate Equations}

To explore electric current in the presence of a field with time-period $T$, we define the time-averaged current $\overline{I}_{\alpha}$ in the lead $\alpha$ as an integral over the time interval $T$, \cite{shot_noise_1,shot_noise_2}   
\begin{align}
\label{elec_current0}
\overline{I}_{\alpha}=\frac{1}{T}\int_{0}^{T}\langle I_{\alpha}(t)\rangle dt.
\end{align}
The symbol $\langle~\rangle$ denotes ensemble average, e.g., $\langle I_{\alpha}(t)\rangle$ is the ensemble-averaged current in the lead $\alpha$ at time $t$ \cite{quantum_transport3} , 
\begin{align}
\label{elec_current}
\langle I_{\alpha}(t)\rangle&=-\textrm{e}\frac{d \langle N_{\alpha}(t)\rangle}{dt} = -\textrm{e}\frac{d \langle\Delta N_{\alpha}(t)\rangle}{dt},
\end{align}
where the ensemble-averaged electron number change $\langle\Delta N_{\alpha}(t)\rangle$ between time $t$ and $t_0$ is
\begin{align}
\label{charge1}
\langle\Delta N_{\alpha}(t)\rangle&=\langle N_{\alpha}(t)\rangle-\langle N_{\alpha}(t_0)\rangle \nonumber \\
                    &=\sum_{q} \langle b^{\dagger}_{\alpha q}(t)b_{\alpha q}(t)\rangle - \langle b^{\dagger}_{\alpha q}(t_0)b_{\alpha q}(t_0)\rangle.
\end{align}

Eq.~(\ref{elec_current}) indicates that as the ensemble-averaged electron number in the lead $\alpha$ decreases with time, i.e., $d \langle N_{\alpha}(t)\rangle/dt < 0$, the current flows into the electrode $\alpha$, i.e., $\langle I_{\alpha}(t) \rangle>0$. At the initial time $t=t_0$ all electrons are at equilibrium, i.e., $\langle a^{\dagger}_{n'}(t_0)a_{n}(t_0)\rangle=\delta_{n'n}\langle N_n(t_0)\rangle$, $\langle b^{\dagger}_{\alpha'q'}(t_0)b_{\alpha q}(t_0)\rangle=\delta_{\alpha'\alpha}\delta_{q'q}f_{\alpha}(\epsilon_{\alpha q})$. In addition, we assume that at $t=t_0$ the electrons between the lead and the molecule have no interaction, i.e., $\langle a^{\dagger}_n(t_0)b_{\alpha q}(t_0)\rangle=\langle b^{\dagger}_{\alpha q}(t_0)a_{n}(t_0)\rangle$=0. By virtue of these conditions, substituting Eq.~(\ref{fundamental}) into Eq.~(\ref{charge1}) results in the relation  
\begin{widetext}
\begin{align}
\label{charge2}
\langle\Delta N_{\alpha}(t)\rangle&=\sum_{qn}|\langle\alpha q|U(t,t_0)|n\rangle|^2\langle N_n(t_0) \rangle 
                      +\sum_{\alpha'qq'}\left(|\langle\alpha q|U(t,t_0)|\alpha' q'\rangle|^2 f_{\alpha'}(\epsilon_{\alpha' q'})-f_{\alpha}(\epsilon_{\alpha q})\right) \\
\label{charge3}
 &=\sum_{qn}\left(|\langle\alpha q|U(t,t_0)|n\rangle|^2\langle N_n(t_0)\rangle-|\langle n|U(t,t_0)|\alpha q\rangle|^2f_{\alpha}(\epsilon_{\alpha q})\right) \nonumber \\
                      &+\sum_{\alpha'\neq\alpha} \sum_{qq'}\left(|\langle\alpha q|U(t,t_0)|\alpha' q'\rangle|^2 f_{\alpha'}(\epsilon_{\alpha' q'})-|\langle\alpha' q'|U(t,t_0)|\alpha q\rangle|^2 f_{\alpha}(\epsilon_{\alpha q})\right).
\end{align}
\end{widetext}
Here the backscattering term has been eliminated by invoking the relation $f_{\alpha}(\epsilon_{\alpha q})=\langle \alpha q|U^{\dagger}(t,t_0)U(t,t_0)|\alpha q\rangle f_{\alpha}(\epsilon_{\alpha q})$ and $1=\sum_{n}|n\rangle\langle n|+\sum_{\alpha q}|\alpha q\rangle \langle \alpha q|$.

After substituting Eq.~(\ref{charge3}) into Eq.~(\ref{elec_current}) and taking the long-time limit $t_0\rightarrow-\infty$, the ensemble-averaged current $\langle I_{\alpha}(t)\rangle$ can be expressed as
\begin{align}
\label{current1}
\langle I_{\alpha}(t)\rangle&=-\textrm{e} \sum_{qn}\left[ k_{\alpha q,n}(t)\langle N_n(t_0) \rangle-k_{n,\alpha q}(t)f_{\alpha}(\epsilon_{\alpha q})\right] \nonumber \\
                         &- \textrm{e} \sum_{\alpha'\neq\alpha}\sum_{qq'}\left[ k_{\alpha q,\alpha' q'}(t)f_{\alpha'}(\epsilon_{\alpha' q'})-k_{\alpha' q',\alpha q}(t)f_{\alpha}(\epsilon_{\alpha q})\right], 
\end{align} 
where $k_{n, \alpha q}(t)\equiv\lim_{t_0\rightarrow-\infty}\frac{d}{dt}|\langle n|U(t,t_0)|\alpha q\rangle|^2$ denotes the rate that an electron tunnels from the orbital $|\alpha q\rangle$ in the lead $\alpha$ to the orbital $|n\rangle$ on the molecule, and $k_{\alpha q,\alpha' q'}(t)\equiv\lim_{t_0\rightarrow-\infty}\frac{d}{dt}|\langle \alpha q|U(t,t_0)|\alpha' q'\rangle|^2$ denotes the rate that an electron tunnels from the orbital $|\alpha' q'\rangle$ in the lead $\alpha'$ to the orbital $|\alpha q\rangle$ in the lead $\alpha$. In the long-time limit $t_0\rightarrow-\infty$, all transient currents die out so $k_{\alpha q,n}\langle N_n(t_0) \rangle$ can be ignored in our analysis. In addition, $k_{n,\alpha q}f_{\alpha}(\epsilon_{\alpha q})$ corresponds to a periodic charging of the molecule driven by external time-periodic fields \cite{shot_noise_1,shot_noise_2} and it contributes zero current over a period $T$ (The detailed derivation is in Appendix \ref{appendix1}). Therefore, the time-averaged current in lead $\alpha$ reads   
\begin{align}
\label{elec_current01}
\overline{I}_{\alpha}&=-\frac{e}{T}\int^{T}_{0}dt \sum_{\alpha'\neq\alpha}^{\kappa}\sum_{qq'}[k_{\alpha q,\alpha' q'}(t)f_{\alpha'}(\epsilon_{\alpha' q'}) \nonumber \\ 
&-k_{\alpha' q',\alpha q}(t)f_{\alpha}(\epsilon_{\alpha q})],
\end{align}
indicating that the current is only relevant to the lead-to-lead tunneling rates and the Fermi functions.

Furthermore, using the following relations,
\begin{align}
\frac{d\langle\alpha'q'|U^{\dagger}(t,t_0)|\alpha q\rangle}{dt}&=\frac{i}{\hbar}[\epsilon_{\alpha q}\langle\alpha'q'|U^{\dagger}(t,t_0)|\alpha q\rangle \nonumber \\
&+\sum_n V_{\alpha q,n}^{*}\langle\alpha'q'|U^{\dagger}(t,t_0)|n\rangle], \\
\frac{d\langle\alpha q|U(t,t_0)|\alpha' q'\rangle}{dt}&=-\frac{i}{\hbar}[\epsilon_{\alpha q}\langle\alpha q|U(t,t_0)|\alpha' q'\rangle \nonumber \\
&+\sum_n V_{\alpha q,n}\langle n|U(t,t_0)|\alpha'q'\rangle],
\end{align}
the lead-to-lead tunneling rate $k_{\alpha q,\alpha' q'}(t)$ can be expressed in terms of $\langle \alpha q|U(t,t_0)|n\rangle$ and $\langle \alpha q|U(t,t_0)|\alpha' q'\rangle$ as
\begin{align}
\label{rate1}
k_{\alpha q,\alpha' q'}(t)&=\lim_{t_0 \rightarrow -\infty}\frac{d}{dt}|\langle\alpha q|U(t,t_0)|\alpha'q'\rangle|^2 \nonumber \\
&=\lim_{t_0 \rightarrow -\infty}\frac{i}{\hbar}\sum_n V^*_{\alpha q,n} \langle\alpha' q'|U^{\dagger}(t,t_0)|n\rangle \nonumber \\
&\times\langle\alpha q|U(t,t_0)|\alpha' q'\rangle +c.c.
\end{align}

\subsection{Propagators in terms of Green's functions}

To derive the ensemble-averaged current $\langle I_{\alpha}(t)\rangle$, the evaluation of the matrix elements of the propagators, e.g., $\langle \alpha q|U(t,t_0)|n\rangle$ and $\langle \alpha q|U(t,t_0)|\alpha' q'\rangle$, is needed. We start from the interaction picture and separate the total Hamiltonian into the uncoupled Hamiltonian $H_{0}(t)=H_{\textrm{mol}}(t)+H_{\textrm{lead}}$ and the coupling Hamiltonian $H_{\textrm{int}}=H_{\textrm{coup}}$. The propagator $U_0(t,t_0)$ associated with $H_{0}(t)$ is governed by the equation
\begin{align}
\label{evolution2}
i\hbar\frac{d U_0(t,t_0)}{d t}=H_{0}(t)U_0(t,t_0).
\end{align} 
It is readily seen that $U_0(t,t_0)=U_{\textrm{mol}}(t,t_0)U_{\textrm{lead}}(t,t_0)$, where $U_{\textrm{lead}}(t,t_0)=\exp(-\frac{i}{\hbar}H_{\textrm{lead}}(t-t_0))$ and $U_{\textrm{mol}}(t,t_0)=T_+\exp(-\frac{i}{\hbar}\int_{t_0}^{t} H_{\textrm{mol}}(t') dt')$. Here, $T_+$ is the time ordering operator.

By virtue of Eqs.~(\ref{evolution2_1}) and (\ref{evolution2}), we can obtain
\begin{align}
\label{evolution3}
U(t,t_0)=U_0(t,t_0)-\frac{i}{\hbar}\int^{t}_{t_0}dt'U_0(t,t')H_{\textrm{coup}}U(t',t_0).
\end{align}

By substituting Eq.~(\ref{evolution3}) into $\langle \alpha q|U(t,t_0)|n \rangle$, $\langle n|U(t,t_0)|\alpha q \rangle$ and $\langle \alpha q|U(t,t_0)|\alpha' q' \rangle$, and using the fact that $\langle \alpha q|U_0(t,t_0)|n\rangle=0$ and $\langle \alpha q|U_0(t,t')|\alpha' q'\rangle=\delta_{\alpha\alpha'}\delta_{qq'}\exp(-\frac{i}{\hbar}\epsilon_{\alpha q}(t-t'))$, we find
\begin{widetext}
\begin{align}
\label{evolution6}
\langle n|U(t,t_0)|\alpha q \rangle &= -\frac{i}{\hbar}\sum_{n'}V_{\alpha q,n'}^* \int^{t}_{t_0}dt' e^{-\frac{i}{\hbar}\epsilon_{\alpha q}(t'-t_0)}\langle n|U(t,t')|n' \rangle, \\
\label{evolution5x}
\langle \alpha q|U(t,t_0)|\alpha' q' \rangle&=\delta_{\alpha\alpha'}\delta_{qq'}e^{-\frac{i}{\hbar}\epsilon_{\alpha q}(t-t_0)} 
																					-\frac{1}{\hbar^2}\sum_{n}\sum_{n'}V_{\alpha q,n}V_{\alpha' q',n'}^*\int^{t}_{t_0}dt'e^{-\frac{i}{\hbar}\epsilon_{\alpha q}(t-t')} \nonumber\\
																					&\times\int^{t'}_{t}dt''e^{-\frac{i}{\hbar}\epsilon_{\alpha' q'}(t''-t_0)}\langle n|U(t',t'')|n'\rangle. 
\end{align}
\end{widetext}

Eq.~(\ref{evolution6}) and Eq.~(\ref{evolution5x}) can be further simplified using the retarded and advanced Green's functions defined as
\begin{align}
\label{green1}
G^{\textrm{R}}(t,t')&\equiv-i\theta(t-t')U(t,t'), \\
\label{green1_1} 
G^{\textrm{A}}(t,t')&\equiv i\theta(t'-t)U(t,t'),
\end{align}
where $\theta(t-t')$ is the Heaviside function and Eq.~(\ref{green1}) satisfies $G^{\textrm{R}}(t,t')=[G^{\textrm{A}}(t,t')]^{\dagger}$ and $U(t,t')=i\hbar(G^{\textrm{R}}(t,t')-G^{\textrm{A}}(t,t'))$. Note that $G^{\textrm{R}}(t,t')$ and $G^{\textrm{A}}(t,t')$ are the Green's functions for the total system. From Eq.~(\ref{green1}), the Fourier transforms of the retarded and advanced Green's functions in the energy domain are, respectively,
\begin{align}
\label{green2_1}
G^{\textrm{R}}(t,\epsilon)&\equiv\lim_{\eta\rightarrow+0}G(t,\epsilon+i\eta)  \nonumber \\
&=-\frac{i}{\hbar}\lim_{\eta\rightarrow+0}\int_{0}^{\infty}d\tau e^{\frac{i(\epsilon+i\eta)\tau}{\hbar}}U(t,t-\tau), 
\end{align}
\begin{align}
\label{green2_2}
G^{\textrm{A}}(t,\epsilon)&\equiv\lim_{\eta\rightarrow+0}G(t,\epsilon-i\eta) \nonumber \\
&=\frac{i}{\hbar}\lim_{\eta\rightarrow+0}\int_{-\infty}^{0}d\tau e^{\frac{i(\epsilon-i\eta)\tau}{\hbar}}U(t,t-\tau),
\end{align}
in which $G^{\textrm{R}}(t,\epsilon)=[G^{\textrm{A}}(t,\epsilon)]^{\dagger}$. Note that the total Hamiltonian has time-periodic symmetry $H_{\textrm{tot}}(t)=H_{\textrm{tot}}(t+T)$, leading to $U(t,t')=U(t+T,t'+T)$ and $G^{\textrm{R(A)}}(t,\epsilon)=G^{\textrm{R(A)}}(t+T,\epsilon)$.
As a result, the retarded and advanced Green's function in the energy domain can be expanded in a Fourier series
\begin{align}
\label{green_fourier}
G^R(t,\epsilon)=\sum^{\infty}_{k=-\infty}G^{\textrm{R}(k)}(\epsilon)e^{-ik\omega t},
\end{align}
where the Fourier coefficient $G^{\textrm{R}(k)}(\epsilon)$ is
\begin{align}
\label{green_fourier2}
G^{\textrm{R}(k)}(\epsilon)=\frac{1}{T}\int^{T}_{0}G^{\textrm{R}}(t,\epsilon)e^{ik\omega t} dt.
\end{align}

It can be readily shown that by letting $t_0\rightarrow-\infty$, $\tau=t-t'$, and $d\tau=-dt'$ in Eq.~(\ref{evolution6}), and making use of Eq.~(\ref{green2_1}), Eq.~(\ref{evolution6}) becomes
\begin{align}
\label{rate_green2}
\langle n|U(t,t_0)|\alpha q \rangle &= e^{-\frac{i}{\hbar}\epsilon_{\alpha q}(t-t_0)}\sum_{n'}V_{\alpha q,n'}^*G^{\textrm{R}}_{nn'}(t,\epsilon_{\alpha q}),
\end{align}
where $G^{\textrm{R}}_{nn'}(t,\epsilon_{\alpha q})$ stands for $\langle n|G^{\textrm{R}}(t,\epsilon_{\alpha q})|n'\rangle$ for convenience. 

Similarly, by letting $t_0\rightarrow-\infty$, $\tau=t'-t''$ and $d\tau=-dt''$, Eq.~(\ref{evolution5x}) becomes
\begin{widetext}
\begin{align}
\label{rate_green3}
\langle \alpha q|U(t,t_0)|\alpha' q' \rangle = -\frac{i}{\hbar}e^{-\frac{i}{\hbar}(\epsilon_{\alpha q}t-\epsilon_{\alpha' q'}t_0)}\sum_{n}\sum_{n'}V_{\alpha q,n}V_{\alpha' q',n'}^* 
                                             \int^{t}_{-\infty}dt' e^{\frac{i}{\hbar}(\epsilon_{\alpha q}-\epsilon_{\alpha' q'})t'} G^{\textrm{R}}_{nn'}(t',\epsilon_{\alpha' q'}).
\end{align}
Furthermore, by letting $\tau=t-t'$ and $d\tau=-dt'$ in Eq.~(\ref{rate_green3}), the matrix element $\langle \alpha q|U(t,t_0)|\alpha' q' \rangle$ becomes
\begin{align}
\label{rate_green4}
\langle \alpha q|U(t,t_0)|\alpha' q' \rangle = -\frac{i}{\hbar}e^{-\frac{i}{\hbar}\epsilon_{\alpha' q'}(t-t_0)}\sum_{n}\sum_{n'}V_{\alpha q,n}V_{\alpha' q',n'}^* 
                                              \int^{\infty}_{0}d\tau e^{-\frac{i}{\hbar}(\epsilon_{\alpha q}-\epsilon_{\alpha' q'})\tau} G^{\textrm{R}}_{nn'}(t-\tau,\epsilon_{\alpha' q'}).
\end{align}
\end{widetext}
Note that we do not show the term $\delta_{\alpha\alpha'}\delta_{qq'}e^{-\frac{i}{\hbar}\epsilon_{\alpha q}(t-t_0)}$ in Eq.~(\ref{rate_green3}) and Eq.~(\ref{rate_green4}) because of the condition $\alpha\neq\alpha'$ in Eq.~(\ref{elec_current01}). Eq.~(\ref{rate_green2}) and Eq.~(\ref{rate_green4}) can be used for expressing $k_{\alpha' q',\alpha q}(t)$ and $k_{\alpha q,\alpha' q'}(t)$ in Eq.~(\ref{elec_current01}) in terms of the retarded (advanced) Green's functions.

\subsection{Transmission and Landauer-type Formula}

In most of the literature related to coherent quantum transport, the current formula is expressed in terms of transmission functions. In this section, we will show that the rate equation (Eq.~(\ref{elec_current01})) and the lead-to-lead tunneling rates correspond to the Landauer-type formula and transmission functions, respectively.    

Use of $\langle \alpha'q'|U^{\dagger}(t,t_0)|n\rangle=(\langle n|U(t,t_0)|\alpha'q'\rangle)^*$ and substitution of Eq.~(\ref{rate_green2}) and Eq.~(\ref{rate_green4}) into Eq.~(\ref{rate1}) give
\begin{widetext}
\begin{align}
\label{rate3}
k_{\alpha q,\alpha'q'}(t)&=\lim_{t_0\rightarrow-\infty}\frac{d}{dt}|\langle\alpha q|U(t,t_0)|\alpha'q'\rangle|^2 \nonumber \\ 
&= \frac{1}{\hbar^2}\sum_{n_1n_2n_3n_4}V^*_{\alpha q,n_1}V_{\alpha'q',n_2}V_{\alpha q,n_3}V^*_{\alpha'q',n_4} 
 \int^{\infty}_{0}d\tau e^{-\frac{i}{\hbar}(\epsilon_{\alpha q}-\epsilon_{\alpha'q'})\tau} 
[G^{\textrm{R}}_{n_1n_2}(t,\epsilon_{\alpha' q'})]^*G^{\textrm{R}}_{n_3n_4}(t-\tau,\epsilon_{\alpha' q'})
+c.c.,
\end{align}
\end{widetext}
where we use $n_1$ and $n_2$ instead of $n$ and $n'$ in Eq.~(\ref{rate_green2}) as well as $n_3$ and $n_4$ instead of $n$ and $n'$ in Eq.~(\ref{rate_green4}).
By invoking Eq.~(\ref{rate3}) and after some manipulations, $\sum_{qq'}k_{\alpha q,\alpha'q'}(t)f(\epsilon_{\alpha'q'})$ can be written as 
\begin{widetext}
\begin{align}
\label{rate4}
\sum_{qq'}k_{\alpha q,\alpha'q'}(t)f_{\alpha'}(\epsilon_{\alpha'q'})&=\frac{1}{\hbar}\frac{1}{2\pi}\sum_{n_1n_2n_3n_4}\int^{\infty}_{0}d\tau \left(\frac{1}{\hbar}\sum_{q}V^*_{\alpha q,n_1}V_{\alpha q,n_3}e^{-\frac{i}{\hbar}\epsilon_{\alpha q}\tau}\right)   \nonumber \\         &\times 2\pi\sum_{q'}V_{\alpha'q',n_2}V^*_{\alpha'q',n_4}\exp(\frac{i}{\hbar}\epsilon_{\alpha' q'}\tau)  
                                                       [G^{\textrm{R}}_{n_1n_2}(t,\epsilon_{\alpha' q'})]^*G^{\textrm{R}}_{n_3n_4}(t-\tau,\epsilon_{\alpha' q'}) f(\epsilon_{\alpha'q'}) 
+c.c.
\end{align}
\end{widetext}

By using Eq.~(\ref{self1}), Eq.~(\ref{self2}) and $\int\Gamma_{\alpha,n n'}(\epsilon)g(\epsilon_{\alpha q}) d\epsilon=2\pi\sum_{q}V^{*}_{\alpha q,n}V_{\alpha q,n'}g(\epsilon)$, Eq.~(\ref{rate4}) becomes
\begin{align}
\label{rate5}
&\sum_{qq'}k_{\alpha q,\alpha'q'}f_{\alpha'}(\epsilon_{\alpha'q'}) = \nonumber \\
&\frac{1}{h}\sum_{n_1n_2n_3n_4}\int^{\infty}_{0}d\tau \Gamma_{\alpha,n_1 n_3}(\tau)\int^{\infty}_{-\infty}d\epsilon e^{\frac{i}{\hbar}\epsilon\tau} \Gamma_{\alpha',n_4 n_2}(\epsilon)    \nonumber \\         
                                                       &\times[G^{\textrm{R}}_{n_1n_2}(t,\epsilon)]^*G^{\textrm{R}}_{n_3n_4}(t-\tau,\epsilon) f(\epsilon) +c.c.
\end{align}

Then, by use of Eq.~(\ref{green_fourier}), expansion of $[G^{\textrm{R}}_{n_1n_2}(t,\epsilon)]^*$ and $G^{\textrm{R}}_{n_3n_4}(t-\tau,\epsilon)$ in Eq.~(\ref{rate5}) in a Fourier series gives
\begin{align}
\label{rate6}
&\sum_{qq'}k_{\alpha q,\alpha'q'}f_{\alpha'}(\epsilon_{\alpha'q'}) = \nonumber \\
&\frac{1}{h}\sum_{k,k'=-\infty}^{\infty}e^{i(k'-k)\omega t} \sum_{n_1n_2n_3n_4}\int^{\infty}_{-\infty}d\epsilon \int^{\infty}_{0}d\tau e^{\frac{i(\epsilon+k\hbar\omega)\tau}{\hbar}} 
 \nonumber \\ 
                   &\times\Gamma_{\alpha,n_1 n_3}(\tau)  \Gamma_{\alpha',n_4 n_2}(\epsilon)[G_{n_1n_2}^{\textrm{R}(k')}(\epsilon)]^*G_{n_3n_4}^{\textrm{R}(k)}(\epsilon) f_{\alpha'}(\epsilon)+c.c.
\end{align}

Since the complex conjugate part in Eq.~(\ref{rate6}) contributes $\int^{0}_{-\infty}d\tau$, we can make use of the Fourier transform of the coupling function, i.e., $\Gamma_{\alpha,n_1 n_3}(\epsilon+k\hbar\omega)=\int_{-\infty}^{\infty}d\tau e^{\frac{i(\epsilon+k\hbar\omega)\tau}{\hbar}}\Gamma_{\alpha,n_1 n_3}(\tau)$, and reduce Eq.~(\ref{rate6}) to  
\begin{align}
\label{rate7}
&\sum_{qq'}k_{\alpha q,\alpha'q'}f_{\alpha'}(\epsilon_{\alpha'q'})=\frac{1}{h}\sum_{k,k'=-\infty}^{\infty}e^{i(k'-k)\omega t}\times \nonumber \\
&\int^{\infty}_{-\infty}d\epsilon \textrm{Tr}[\Gamma_{\alpha}(\epsilon+k\hbar\omega)G^{\textrm{R}(k)}(\epsilon)\Gamma_{\alpha'}(\epsilon)G^{\textrm{A}(k')}(\epsilon)] f_{\alpha'}(\epsilon), 
\end{align}
where $\textrm{Tr}[\Gamma_{\alpha}(\epsilon+k\hbar\omega)G^{\textrm{R}(k)}(\epsilon)\Gamma_{\alpha'}(\epsilon)G^{\textrm{A}(k')}(\epsilon)]=\sum_{n_1n_2n_3n_4} \Gamma_{\alpha,n_1 n_3}(\epsilon+k\hbar\omega)G_{n_3n_4}^{\textrm{R}(k)}(\epsilon)\Gamma_{\alpha',n_4 n_2}(\epsilon)G_{n_2n_1}^{\textrm{A}(k')}(\epsilon)$. $\sum_{qq'}k_{\alpha' q',\alpha q}f_{\alpha}(\epsilon_{\alpha q})$ can be derived by repeating the procedure from Eq.~(\ref{rate4}) to Eq.~(\ref{rate7}).

Finally, substitution of Eq.~(\ref{rate7}) into Eq.~(\ref{elec_current01}) and use of $\frac{1}{T}\int^{T}_{0}dt e^{i(k'-k)\omega t}=\delta(k'-k)$ give the time-averaged current without spin degeneracy
\begin{align}
\label{elec_current02}
\overline{I}_{\alpha}=\frac{\textrm{e}}{h}\sum_{\alpha'\neq\alpha}^{\kappa}\sum_{k=-\infty}^{\infty}\int^{\infty}_{-\infty}d\epsilon [T^{(k)}_{\alpha' \alpha}(\epsilon)f_{\alpha}(\epsilon)-T^{(k)}_{\alpha \alpha'}(\epsilon)f_{\alpha'}(\epsilon)],
\end{align}
and the transmission functions
\begin{align}
\label{transmission}
T^{(k)}_{\alpha \alpha'}(\epsilon)&=\textrm{Tr}[\Gamma_{\alpha}(\epsilon+k\hbar\omega)G^{\textrm{R}(k)}(\epsilon)\Gamma_{\alpha'}(\epsilon)G^{\textrm{A}(k)}(\epsilon)],
\end{align}
which corresponds to the tunneling of an electron from the lead $\alpha'$ to the lead $\alpha$ with energy $\epsilon$ accompanied by k-photon absorption ($k>0$) or emission ($k<0$). Note that in general cases $T^{(k)}_{\alpha \alpha'}(\epsilon)\neq T^{(k)}_{\alpha' \alpha}(\epsilon)$ if molecules have no generalized parity symmetry \cite{rectifier,review_floquet}.

Eq.~(\ref{elec_current02}) enables dealing with transport in multi-terminal systems in a time-periodic driving field and it is straightforward to show that $\sum_{\alpha}\overline{I}_{\alpha}=0$ which satisfies the continuity equation. The correspondence between Eq.~(\ref{elec_current01}) and  Eq.~(\ref{elec_current02}) reveals the connection between the lead-to-lead tunneling rates and the transmission functions. In addition, the trace form in Eq.~(\ref{transmission}) enables the formulation to be applied to realistic models, e.g., a molecular Hamiltonians derived from the density-functional method in a maximally localized Wannier function representation \cite{MLWF1,MLWF2}.

In the absence of the external driving field ($H_{\textrm{mol}}(t)$ is time-independent), using the fact that $T^{(k)}_{\alpha \alpha'}(\epsilon) =T^{(k)}_{\alpha' \alpha}(\epsilon)=\delta_{k,0}T^{(k)}_{\alpha \alpha'}(\epsilon)$ reduces Eq.~(\ref{elec_current02}) to the Landauer-type current formula, 
\begin{align}
\label{elec_current03}
\overline{I}_{\alpha}=\frac{\textrm{e}}{h}\sum_{\alpha'\neq\alpha}^{\kappa}\int^{\infty}_{-\infty}d\epsilon T^{(0)}_{\alpha \alpha'}(\epsilon)[f_{\alpha}(\epsilon)-f_{\alpha'}(\epsilon)],
\end{align}
where $T^{(0)}_{\alpha \alpha'}(\epsilon)=\textrm{Tr}[\Gamma_{\alpha}(\epsilon)G^{\textrm{R}(0)}(\epsilon)\Gamma_{\alpha'}(\epsilon)G^{\textrm{A}(0)}(\epsilon)]$, which is consistent with the results derived from Fisher-Lee relation \cite{FisherLee} and Meir-Wingreen formula \cite{Meir}.

For a two terminal system, i.e., $\alpha=$ R and $\alpha'=$ L, assuming that the coupling function satisfies $\Gamma_{\textrm{L}}(\epsilon)=|u\rangle \Gamma_{\textrm{L},uu}(\epsilon) \langle u|$ and $\Gamma_{\textrm{R}}(\epsilon)=|v\rangle \Gamma_{\textrm{R},vv}(\epsilon) \langle v|$, Eq.~(\ref{elec_current02}) reduces to
\begin{align}
\label{elec_current04}
\overline{I}_{\textrm{R}}=\frac{\textrm{e}}{h}\sum_{k=-\infty}^{\infty}\int^{\infty}_{-\infty}d\epsilon [T^{(k)}_{\textrm{LR}}(\epsilon)f_{\textrm{R}}(\epsilon)-T^{(k)}_{\textrm{RL}}(\epsilon)f_{\textrm{L}}(\epsilon)],
\end{align}
where $T^{(k)}_{\textrm{RL}}(\epsilon)=\Gamma_{\textrm{R},vv}(\epsilon+k\hbar\omega)\Gamma_{\textrm{L},uu}(\epsilon)|G^{\textrm{R}(k)}_{vu}(\epsilon)|^2$ and $T^{(k)}_{\textrm{LR}}(\epsilon)=\Gamma_{\textrm{L},uu}(\epsilon+k\hbar\omega)\Gamma_{\textrm{R},vv}(\epsilon)|G^{\textrm{R}(k)}_{uv}(\epsilon)|^2$, which are consistent with the results derived by H\"anggi et al. \cite{shot_noise_1,shot_noise_2}.

\subsection{Solving for $G^{\textrm{R}(k)}_{nn'}(\epsilon)$ Using the Wide Band Limit Approximation and Non-Hermitian Floquet Theory}\label{section2-6}

To solve for the Fourier coefficients of the retarded (advanced) Green's functions, we start from the Schr\"odinger picture and consider the dynamics of $\langle n|U(t,t_0)|n'\rangle$,
\begin{align}
\label{sch_pro2}
i\hbar\frac{d}{dt}\langle n|U(t,t_0)|n'\rangle &=\sum_{n''} [H_{\textrm{mol}}(t)]_{nn''} \langle n''|U(t,t_0)|n' \rangle \nonumber \\
&+ \sum_{\alpha q} V_{\alpha q,n}^* \langle \alpha q|U(t,t_0)|n' \rangle.
\end{align}
The first and second terms on the right-hand side correspond to the dynamics governed by the molecular Hamiltonian and influenced by the leads, respectively. By virtue of Eq.~(\ref{self2}) and Eq.~(\ref{evolution3}), the last term in Eq.~(\ref{sch_pro2}) can be expressed in terms of memory functions as 
\begin{align}
\label{sch_pro3}
-i \sum_{\alpha}\sum_{n''} \int^{t}_{t_0}dt'\Gamma_{\alpha,nn''}(t-t')\langle n''|U(t',t_0)|n' \rangle.
\end{align}
Moreover, we adopt the Markov process approximation for the memory function, i.e., $\Gamma_{\alpha,nn''}(t-t')=\overline{{\Gamma}}_{\alpha,nn''}\delta(t-t')$, and the last term in Eq.~(\ref{sch_pro2}) becomes
\begin{align}
\label{sch_pro4}
-\frac{i}{2} \sum_{\alpha}\sum_{n''} \overline{{\Gamma}}_{\alpha,nn''}\langle n''|U(t,t_0)|n' \rangle, 
\end{align}
where the factor of $1/2$ comes from $\int^{t}_{t_0}dt'\delta(t-t')\langle n''|U(t',t_0)|n' \rangle=\langle n''|U(t,t_0)|n' \rangle/2$. Note that the Markov process approximation for the memory function is equivalent to the wide band limit approximation, i.e., $\Gamma_{\alpha,n n'}(\epsilon)=\overline{\Gamma}_{\alpha,n n'}=\textrm{constant}$. As $\Gamma_{\alpha,n n'}(\epsilon)$ is energy-independent, its Fourier transform is $\overline{{\Gamma}}_{\alpha,nn''}\delta(t-t')$.

Substitution of Eq.~(\ref{sch_pro4}) into Eq.~(\ref{sch_pro2}) gives
\begin{align}
\label{sch_pro5}
&i\hbar\frac{d}{dt}\langle n|U(t,t_0)|n'\rangle \nonumber \\
&=\left(\sum_{n''} [H_{\textrm{mol}}(t)]_{nn''} -\frac{i}{2} \sum_{\alpha}\sum_{n''} \overline{{\Gamma}}_{\alpha,nn''} \right)\langle n''|U(t,t_0)|n' \rangle,
\end{align}
which can be equivalently written as 
\begin{align}
\label{sch_pro6}
i\hbar\frac{d}{dt}|\psi(t)\rangle= \left(H_{\textrm{mol}}(t) -\frac{i}{2} \sum_{\alpha} \overline{{\Gamma}}_{\alpha}\right)|\psi(t)\rangle,
\end{align}
where the wavefunction $|\psi(t)\rangle$ is related to the propagator $U(t,t_0)$ via the relation $|\psi(t)\rangle=U(t,t_0)|\psi(t_0)\rangle$. Eq.~(\ref{sch_pro6}) is a linear, time-periodic ($H_{\textrm{mol}}(t)=H_{\textrm{mol}}(t+T)$), non-Hermitian ordinary differential equation, whose complete solution based on the Floquet theory can be written as 
\begin{align}
\label{Floquet1}
|\psi_{\lambda}(t)\rangle = \exp[-i(\epsilon_{\lambda}/\hbar-i\gamma_{\lambda})t]|\phi_{\lambda}(t)\rangle,
\end{align} 
where $\exp[-i(\epsilon_{\lambda}/\hbar-i\gamma_{\lambda})t]$ and $|\phi_{\lambda}(t)\rangle$ are the Floquet exponent (characteristic exponent) and Floquet eigenstates respectively. Moreover, $\epsilon_{\lambda}$ and $1/\gamma_{\lambda}$ correspond to the quasienergy and lifetime of the Floquet eigenstates $|\phi_{\lambda}(t)\rangle$. Here the Floquet eigenstate is a periodic function of time, i.e., $|\phi_{\lambda}(t+T)\rangle=|\phi_{\lambda}(t)\rangle$.

By substitution of Eq.~(\ref{Floquet1}) into Eq.~(\ref{sch_pro6}), the underlying time-dependent Schr\"odinger equation can be cast into the Floquet eigenvalue equation
\begin{align}
\label{Floquet2}
H_F|\phi_{\lambda}(t)\rangle = (\epsilon_{\lambda}-i\hbar\gamma_{\lambda})|\phi_{\lambda}(t)\rangle, \\
H_F=H_{\textrm{mol}}(t)-\frac{i}{2} \sum_{\alpha} \overline{{\Gamma}}_{\alpha}-i\hbar\frac{d}{d t},
\end{align} 
where $H_F$ is the so-called Floquet Hamiltonian in an extended Hilbert space \cite{Sambe}. Note that $H_F$ is non-Hermitian Hamiltonian \cite{nonHerm1,nonHerm2} so its adjoint eigenvalue equation satisfies
\begin{align}
\label{Floquet2_1}
H_F^{\dagger}|\chi_{\lambda}(t)\rangle = (\epsilon_{\lambda}+i\hbar\gamma_{\lambda})|\chi_{\lambda}(t)\rangle.
\end{align} 
The eigenstates $|\phi_{\lambda}(t)\rangle$ and the adjoint states $|\chi_{\lambda}(t)\rangle$ form a complete biorthogonal basis in an extended Hilbert space.

Due to the time-periodic symmetry of the Floquet eigenstates $|\phi_{\lambda}(t+T)\rangle=|\phi_{\lambda}(t)\rangle$, a complete solution can be also constructed as \cite{Shirley},
\begin{align}
\label{Floquet3}
|\psi_{\lambda}(t)\rangle = \exp(-i\frac{q_{\lambda,\zeta}}{\hbar}t)|\phi_{\lambda,\zeta}(t)\rangle,
\end{align}
where $q_{\lambda,\zeta}=\epsilon_{\lambda}-i\hbar\gamma_{\lambda}+\zeta\hbar\omega$ for any integer $\zeta$ and $|\phi_{\lambda,\zeta}(t)\rangle=|\phi_{\lambda,0}(t)\rangle \exp(i\zeta\omega t)$. For consistency, we use the notation $|\phi_{\lambda,0}(t)\rangle$ instead of $|\phi_{\lambda}(t)\rangle$ in the following derivations. The quasienergies $\epsilon_{\lambda,0}$ can be mapped into the first Brillouin zone, $\overline{\epsilon}-\hbar\omega/2<\epsilon_{\lambda,0}\leq\overline{\epsilon}+\hbar\omega/2$, where $\overline{\epsilon}$ is an arbitrary chosen real number. Note that for fixed time $t$, the Floquet states of the first Brillouin zone form a complete set in \textsl{R}, i.e., $\sum_{\lambda\in\textrm{1st BZ}}|\phi_{\lambda,0}(t)\rangle\langle\chi_{\lambda,0}(t)|=1$, where 1st BZ denotes the first Brillouin zone \cite{Hanggi_review}.
By substituting Eq.~(\ref{Floquet3}) into Eq.~(\ref{sch_pro6}), we arrive at the following eigenvalue equation
\begin{align}
\label{Floquet4}
H_F|\phi_{\lambda,\zeta}(t)\rangle = q_{\lambda,\zeta}|\phi_{\lambda,\zeta}(t)\rangle,
\end{align} 
where $|\phi_{\lambda,\zeta}(t)\rangle$ has time-periodic symmetry and can be decomposed into a Fourier series,
\begin{align}
\label{Floquetcoef_2}
|\phi_{\lambda,\zeta}(t)\rangle = \sum_{k=-\infty}^{\infty}|\phi^{k}_{\lambda,\zeta}\rangle\exp(ik\omega t).
\end{align} 
Note that different Floquet states have the following property,
\begin{align}
\label{Floquetcoef_3}
|\phi^{k+k'}_{\lambda,\zeta}\rangle=|\phi^{k}_{\lambda,\zeta-k'}\rangle.
\end{align}

To further facilitate the analysis, we substitute $1=\sum_{n}|n\rangle \langle n|$ and Eq.~(\ref{Floquetcoef_2}) into Eq.~(\ref{Floquet4}), and perform the Fourier transform on both sides of Eq.~(\ref{Floquet4}), i.e., $\int_{0}^{T} dt \exp(-ik'\omega t)/T$, and then derive a time-independent infinite-dimensional eigenvalue matrix equation 
\begin{align}
\label{FM}
\sum_{nk} [\overline{H}_{F}]_{n'k',nk}\phi^{n,k}_{\lambda,\zeta} = q_{\lambda\zeta}\phi^{n',k'}_{\lambda,\zeta}.
\end{align} 
Here, $\phi^{n,k}_{\lambda,\zeta}=\langle n|\phi^{k}_{\lambda,\zeta}\rangle$ and $[\overline{H}_{F}]_{n'k',nk}$ denotes the matrix elements of the time-averaged Floquet Hamiltonian over a period $T$. Moreover, $[\overline{H}_{F}]_{n'k',nk}$ can be cast as
\begin{align}
\label{FM1}
[\overline{H}_{F}]_{n'k',nk}
&= [H^{(k'-k)}_{\textrm{mol}}]_{n'n}-\delta_{k'k}\frac{i}{2}\sum_{\alpha} \overline{{\Gamma}}_{\alpha,n'n} \nonumber \\ &+ \delta_{n'n}\delta_{k'k} k\hbar\omega,
\end{align}
where $H^{(k'-k)}_{\textrm{mol}}$ is the Fourier coefficient of $H_{\textrm{mol}}(t)$ defined as 
\begin{align}
\label{Fcoefficeint}
H^{(k'-k)}_{\textrm{mol}}=\frac{1}{T}\int^{T}_{0}dt H_{\textrm{mol}}(t)e^{-i(k'-k)\omega t}.
\end{align}
Similarly, we can derive the adjoint eigenvalue equation of Eq.~(\ref{FM_adj})
\begin{align}
\label{FM_adj}
\sum_{nk} [\overline{H}^{\dagger}_{F}]_{n'k',nk}\chi^{n,k}_{\lambda,\zeta} = \tilde{q}_{\lambda\zeta}\chi^{n',k'}_{\lambda,\zeta}.
\end{align}
With the help of Eq.~(\ref{Floquet1}), it is straightforward to show that the propagator in Eq.~(\ref{sch_pro2}) can be expressed as   
\begin{align}
\label{evolution_nonHermitian}
U(t,t_0)
=\sum_{\lambda\in\textrm{1st BZ}}e^{-i(\frac{\epsilon_{\lambda}}{\hbar}-i\gamma_{\lambda})(t-t_0)}|\phi_{\lambda,0}(t)\rangle \langle \chi_{\lambda,0}(t_0)|.
\end{align}
By combining Eq.~(\ref{green2_1}), Eq.~(\ref{green_fourier2}) and Eq.~(\ref{evolution_nonHermitian}), and making use of Eq.~(\ref{Floquetcoef_3}), the Fourier coefficients of the retarded Green's function can be written as
\begin{align}
\label{green}
G^{\textrm{R}(k)}_{nn'}(\epsilon)&=\sum_{\lambda\in\textrm{1st BZ}}\sum^{+\infty}_{\zeta=-\infty}\frac{ \langle n|\phi^{-k}_{\lambda,\zeta}\rangle\langle \chi^{0}_{\lambda,\zeta}|n'\rangle}{\epsilon-(\epsilon_{\lambda}-i\hbar\gamma_{\lambda}+\zeta\hbar\omega)} \\
\label{green2}
&=\sum_{\lambda\in\textrm{1st BZ}}\sum^{+\infty}_{\zeta=-\infty}\frac{ \phi^{n,-k}_{\lambda,\zeta} \phi_{\lambda,\zeta}^{n', 0}}{\epsilon-q_{\lambda,\zeta}},
\end{align}
where $\phi_{\lambda,\zeta}^{n', 0}=\langle n'|\phi^{0}_{\lambda,\zeta}\rangle=\langle \chi^{0}_{\lambda,\zeta}|n'\rangle$ \cite{nonHerm1}. Note that Eq.~(\ref{green2}) is used to numerically solve for the retarded Green's functions in this paper.

As $H_{\textrm{mol}}$ is time-independent, $G^{(\textrm{R})(k)}_{nn'}(\epsilon)=0$ for $k\neq 0$. As a result, Eq.~(\ref{green2}) can reduce to
\begin{align}
\label{green_zerofield}
G^{\textrm{R}(0)}_{nn'}(\epsilon)=\sum_{\lambda\in\textrm{1st BZ}}\sum^{+\infty}_{\zeta=-\infty}\frac{ \phi^{n,0}_{\lambda,\zeta} \phi_{\lambda,\zeta}^{n', 0}}{\epsilon-q_{\lambda,\zeta}}, 
\end{align}
which is equivalent to the retarded Green's function  
\begin{align}
\label{green_zerofield2}
G^{\textrm{R}}_{nn'}(\epsilon)=\sum_{\nu}\frac{ \tilde{\phi}^{n}_{\nu} \tilde{\phi}_{\nu}^{n'}}{\epsilon-\tilde{q}_{\nu}}, 
\end{align}
derived from a time-independent Hamiltonian eigenvalue matrix equation
\begin{align}
\label{Hamiltonian_zerofield}
\sum_{n}\langle n'|H_{\textrm{mol}}-\frac{i}{2}\sum_{\alpha}\overline{{\Gamma}}_{\alpha}|n\rangle \tilde{\phi}_{\nu}^{n}= \tilde{q}_{\nu}\tilde{\phi}_{\nu}^{n'}.
\end{align}
For a two-level system it is more convenient to use Eq.~(\ref{green_zerofield2}) instead of Eq.~(\ref{green_zerofield}) since the former can be solved analytically (see section \ref{section4-2}).

To summarize, current through a single-molecule junction under a time-periodic field can be computed numerically according to the following five steps: (i) construct the molecular Hamiltonian $H_{\textrm{mol}}(t)$ and coupling functions $\overline{{\Gamma}}_{\alpha}$, (ii) solve the Floquet eigenvalue equation, Eq.~(\ref{FM}), and then obtain its eigenvalues and eigenvectors, (iii) evaluate the matrix element of the retarded Green's function by substitution of the eigenvalues and eigenvectors into Eq.~(\ref{green2}), (iv) compute the transmission functions by using Eq.~(\ref{transmission}), and (v) compute the current by using Eq.~(\ref{elec_current02}).

\section{Two-Terminal Network System: Single-Molecule Optoelectronic Switch}\label{section3}

\begin{figure}
  \centering
  \includegraphics[width=8cm]{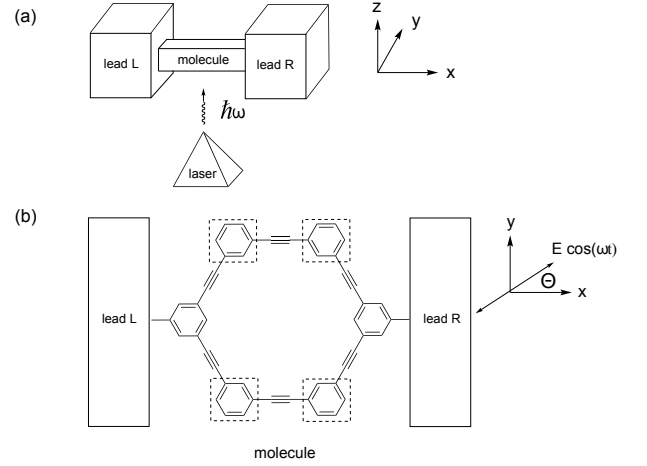}
  \caption{(a) The single-molecule optoelectronic switch is made of two electrodes (left and right leads), PAM, and a monochromatic laser field. (b) The xy-plane projection of the single-molecule optoelectronic device. The PAM molecule is placed in the xy plane of the device and contains four identical meta-benzene based building blocks, indicated by the dashed lines, $\Theta$ is the angle between the polarization direction of the laser and the x axis.}
  \label{fig1}
\end{figure}

Here and in section \ref{section4}, we explore laser-driven transport through three systems with multiple pathways by employing the method derived in the section \ref{section2}. The first system shown in Fig.~\ref{fig1} (a) is a two-terminal molecular junction based on a PAM molecule \cite{PAM1,PAM2} in the presence of a monochromatic laser field. According to a previous study \cite{optoelec}, the first system is an effective optoelectronic switching device due to its large on-off current ratios and weak-field operating conditions, both of which are not sensitive to the Fermi level of the electrodes. However, except for the Fermi level of the electrodes, the previous study did not consider other experimental conditions, e.g., asymmetric molecule-lead coupling and laser polarization, which may affect the large on-off current ratios and weak-field operation. To examine whether the first system is a robust optoelectronic switch, we study laser-driven transport through the PAM molecule by changing  the field amplitude, coupling function strength, and laser polarization direction.

For simplicity, we adopt the H\"uckel model and the electric dipole approximation to describe the molecular Hamiltonian $H_{\textrm{mol}}(t)$ because PAM is a typical conjugated molecule and we only consider weak-field operation. As a result, the molecular Hamiltonian can be expressed as
\begin{align}
\label{molHamiltonian}
H_{\textrm{mol}}(t)&=\sum_{n}(E_{0}-\textrm{e}\mathbf{r}_{n}\cdot\mathbf{E}(t)) a^{\dagger}_{n}a_{n}+\sum_{nn'}\Delta a^{\dagger}_{n}a_{n'}, \\
\label{polarization}
\mathbf{E}(t)&=(\textrm{E}\cos \Theta\hat{x}+\textrm{E}\sin \Theta\hat{y})\cos(\omega t),
\end{align}
where $E_{0}$ is the energy of the $p_z$-orbital on the carbon atoms, $\Delta$ is the resonance integral between directly-bonded carbon atoms, $a^{\dagger}_{n}$ and $a_{n}$ are Fermion operators which create and annihilate an electron in the $p_{z}$-orbital $|n\rangle$ on the n-th carbon of PAM at position $\mathbf{r}_{n}=x_{n}\hat{x}+y_{n}\hat{y}$, and $\mathbf{E}(t)$ is a time-dependent electric field propagating along the z direction with the field amplitude E, frequency $\omega$, and polarization angle $\Theta$ between the laser field and the x axis (see Fig.~\ref{fig1} (b)). Note that $\mathbf{r}_{n}$ is derived from the geometry optimization of PAM at the B3LYP/6-31 G(d) level using the Gaussian 09 program \cite{Gaussian}. The parameters $E_{0}=-6.553$ eV and $\Delta=-2.734$ eV are from photo-electron spectroscopy experiments \cite{HuckelP}. Within the wide band limit approximation, the coupling functions have the form
\begin{align}
\label{Coupmatrix1_1}
 \overline{\Gamma}_{L}=|u\rangle \overline{\Gamma}_{\textrm{L},uu}\langle u|, \\
\label{Coupmatrix1_2}
\overline{\Gamma}_{\textrm{R}}=|v\rangle \overline{\Gamma}_{\textrm{R},vv}\langle v|,
\end{align}
where $|u\rangle$ and $|v\rangle$ respectively denote the $p_z$-orbital on the contact carbon atoms $u$ and $v$. Substituting Eq.~(\ref{molHamiltonian}) into Eq.~(\ref{Fcoefficeint}) gives 
\begin{align}
\label{molHamiltonian2}
H^{(k'-k)}_{\textrm{mol}}&=\delta_{kk'}(\sum_{n}E_{0}a^{\dagger}_{n}a_{n}+\sum_{nn'}\Delta a^{\dagger}_{n}a_{n'})  \nonumber \\
&-\delta_{k k'\pm1}\frac{\textrm{e}\textrm{E}}{2}(x_{n}\cos\Theta+y_{n}\sin\Theta) a^{\dagger}_{n}a_{n}.
\end{align}
Substitution of Eqs.~(\ref{Coupmatrix1_1}), (\ref{Coupmatrix1_2}), and (\ref{molHamiltonian2}) into Eq.~(\ref{FM1}) yields the following time-averaged Floquet Hamiltonian matrix  

\begin{widetext}
\begin{align} 
\label{supermatrix}
\overline{H}_{F}= \bordermatrix{    &          &  k=-2                          & k =-1                      & k=0       & k=1                        & k=2  &   \cr
                                    &  \ddots  &   \vdots                       &    \vdots                  &  \vdots   &  \vdots                    &\vdots&   \cr  
                             k'=-2  &  \cdots  &   A-2\hbar\omega\textbf{I}     &   B                        &   0       &  0                         &  0   & \cdots \cr  
                             k'=-1  &  \cdots  &   B                            &   A-\hbar\omega\textbf{I}  &   B       &  0                         &  0   & \cdots \cr
			                       k'=0   &  \cdots  &   0                            &   B                        &   A       &  B                         &  0   & \cdots \cr 
			                       k'=1   &  \cdots  &   0                            &   0                        &   B       &  A+\hbar\omega\textbf{I}   &  B   & \cdots \cr
				                     k'=2   &  \cdots  &   0                            &   0                        &   0       &  B                         &  A+2\hbar\omega\textbf{I}   &\cdots   \cr
				                            &          &  \vdots                        &   \vdots                   &  \vdots   &  \vdots                    &\vdots& \ddots \cr},
\end{align}
\end{widetext}
where $A$ and $B$ are block matrices with elements $A_{n',n}=[H^{(0)}_{\textrm{mol}}]_{n',n}-(i/2)(\delta_{n'u}\delta_{nu}\overline{\Gamma}_{\textrm{L},uu}+\delta_{n'v}\delta_{nv}\overline{\Gamma}_{\textrm{R},vv})$ and $B_{n',n}=[H^{(\pm 1)}_{\textrm{mol}}]_{n',n}$, respectively. By solving Eq.~(\ref{FM}) and performing steps (iii) -- (v), we can obtain the transmission and the light-driven current through the PAM-based optoelectronic switch.


\subsection{Zero-Field : Destructive Quantum Interference}\label{section3-1}

In an experiment, the measured current through a single-molecule junction is dependent on the chemical nature of the electrodes and linker groups, e.g., the types of electrodes such as Au and Ag \cite{electrode_type1}, electrode conformations such as Au(111) and Au(100), hollow and on-top contacts \cite{hollow1,hollow2,hollow3}, and the types of linker groups such as thiol ($-$SH) and amine ($-\textrm{NH}_{2}$) groups \cite{CPAFM,anchor1}. The influence of these effects can be modeled as the self-energy derived from scattering formulations or non-equilibrium Green's functions, if the system is in the coherent tunneling regime, i.e., molecules with short length and large injection gap at low temperature \cite{molecular_electronics4_1,tunnelingtime1,tunnelingtime2}. Therefore, in this section, we model the influence of the electrodes and the linker groups via changing the values of the coupling functions (the imaginary term of the self-energy) in order to examine whether the destructive quantum interference caused by the PAM-based molecular network is sensitive to molecule-lead couplings.

\begin{figure}
  \centering
  \includegraphics[width=8cm]{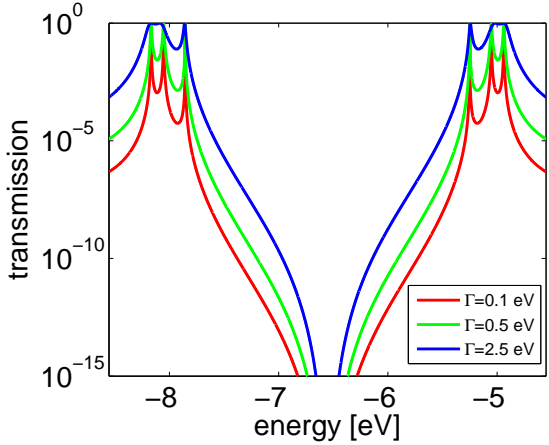}
  \caption{Transmission function of a two-terminal PAM switch in the absence of a laser field with different molecule-lead coupling strength, $\overline{\Gamma}_{\textrm{L},uu}=\overline{\Gamma}_{\textrm{R},vv}=\Gamma$, where the abscissa is the energy of the tunneling electron.}
  \label{fig2}
\end{figure}

Fig.~\ref{fig2} depicts the transmission of the two-terminal PAM junction with different molecule-lead coupling strength in the absence of a laser field. For $\Gamma=0.1$, $0.5$, $2.5$ eV (We consider symmetric molecule-lead couplings, i.e., $\overline{\Gamma}_{\textrm{L},uu}=\overline{\Gamma}_{\textrm{R},vv}=\Gamma$), the transmission is strongly suppressed at energies $\epsilon=E_{0}+\Delta,E_{0},E_{0}-\Delta$ \cite{Interference4,benzene}. Thus, the transmission characteristics are not sensitive to the values of the coupling functions. The transmission suppression originates from destructive quantum interference resulting from the meta-connected benzene unit, indicated by the dashed lines in Fig.~\ref{fig1} (b), and the repeated meta-connected benzene units can suppress the transmission of a tunneling electron and broaden the range of anti-resonance \cite{optoelec}. The values of the coupling functions do not strongly influence the transmission characteristics because the meta-connected benzene units are not directly coupled to the two leads. In addition, the wide anti-resonance range indicates that the current is minute in the small source-drain voltage limit, e.g., $\overline{I}\approx eV_{\textrm{SD}}T/h=1.94\times10^{-16}$ ampere at $T=10^{-10}$ and $V_{\textrm{SD}}=0.05$V. Note that we do not include the spin degeneracy for the computed current. As a result, a two-terminal PAM junction in the absence of a laser field can function as the ``off-state'' of a single-molecule switch because its extremely small transmission is robust to the change of molecule-lead couplings and the Fermi level of the electrode.

 All transmission peaks in Fig.~\ref{fig2} occur in pairs with the center $\epsilon=E_{0}$ because PAM is an alternant hydrocarbon \cite{Salem}. In addition, the molecular orbital energies of PAM are $E_{\text{MO}}=-8.167$ eV, $-8.053$ eV, $-7.855$ eV, $-5.251$ eV, $-5.053$ eV, and $-4.939$ eV, which are in good agreement with the six transmission peaks in Fig.~\ref{fig2}. We do not consider the real part of the self-energy in our analysis because it just shifts the energy of resonant states and does not influence the transmission characteristics.

\subsection{Weak-Field: Photon-Assisted Tunneling}\label{section3-2}

\begin{figure}
  \centering
  \includegraphics[width=8cm]{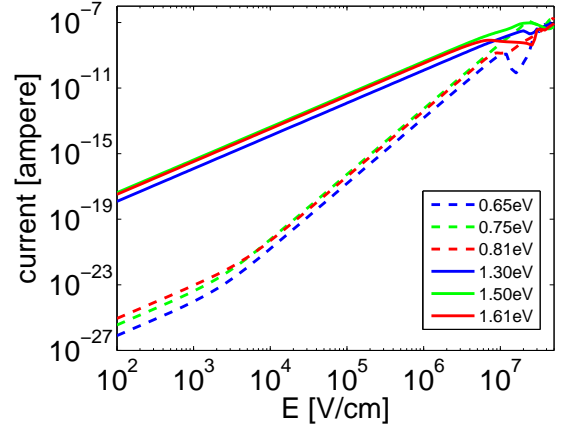}
  \caption{Current-field intensity characteristics of a two-terminal PAM junction for $V_{\textrm{SD}}=0.05$ V, $k_B\theta=5\times 10^{-4}$ eV, $\Gamma=0.05$, $\Theta=0$, $\mu_{\textrm{L}}=\alpha-\textrm{e}V_{\textrm{SD}}/2$, and $\mu_{\textrm{R}}=\alpha+\textrm{e}V_{\textrm{SD}}/2$.}
  \label{fig3}
\end{figure}

Consider the following experimental conditions for a two-terminal PAM junction. We assume a small source-drain voltage ($V_{\textrm{SD}}=0.05$V), low temperature limit ($k_B\theta=5\times 10^{-4}$ eV), and symmetric chemical potentials ($\mu_{\textrm{L}}=\alpha-\textrm{e}V_{\textrm{SD}}/2$ and $\mu_{\textrm{R}}=\alpha+\textrm{e}V_{\textrm{SD}}/2$). Fig.~\ref{fig3} shows that the current-field intensity characteristics of a two-terminal PAM junction. The solid lines are proportional to the second power of the field amplitude $\textrm{E}$ in the range $\textrm{E}<2\times10^6$~V/cm and the dashed lines are proportional to the fourth power of $\textrm{E}$ in the range from $10^4$ to $2\times10^6$~V/cm. This phenomenon is due to photon-assisted tunneling, which occurs under the condition that an integer multiple of laser frequencies are compatible with the energy difference between the resonant-state energies of a two-terminal PAM junction and the Fermi level of electrodes, i.e., $|k|\hbar\omega=|E_{\text{MO}}-\alpha|$. Consider the case of one-photon assisted tunneling ($|k|=1$), the frequency $\hbar\omega$ is $|-4.939+6.553|$ eV $=1.614$eV, consistent with the red solid line in Fig.~\ref{fig3}. As a result, we can conclude that the solid lines and the dashed lines correspond to current induced by one-photon assisted tunneling and two-photon assisted tunneling, respectively.

The field-amplitude power laws of current induced by photon assisted tunneling can be understood by using time-independent, non-Hermitian perturbation theory \cite{nonHerm1} in Appendix \ref{appendix2}. As the electric dipole term in $H_{\textrm{mol}}(t)$ is sufficiently weak, Appendix \ref{appendix2} shows that the one-photon Green's function $G^{(\pm 1)}_{vu}(\epsilon)$ and the two-photon Green's function $G^{(\pm 2)}_{vu}(\epsilon)$ are proportional to $\textrm{E}$ and $\textrm{E}^2$, respectively. The power laws do not hold for $\textrm{E} \gtrsim 2\times10^6$~V/cm because of Stark shifting of the quasi-states. Moreover, we can observe the deviation of the dashed lines from the power laws for a very weak field, i.e., $\textrm{E} \lesssim 10^4$~V/cm, which results from the overlap of one-photon and two-photon assisted tunneling. Therefore, by use of photon-assisted tunneling, a two-terminal PAM junction can function as the ``on-state'' of a single-molecule switch. Moreover, a PAM-based switch may be used for a wide range of applications due to its weak-field operating conditions. Fig.~\ref{fig3} shows the effective range of the field amplitude power laws for the PAM-based optoelectronic switch, indicating that the power laws may be applied to examine the one-photon and two-photon assisted tunneling in a molecular junction experiment.

\subsection{Asymmetric Coupling}\label{section3-3}

Although great progress has been made in fabrication techniques over the past thirty years, it is still not easy to experimentally create a high-quality symmetric molecular junction. The effects of asymmetric molecular junctions on electron transport in the presence of a driving field are non-trivial since they can lead to the coherent quantum ratchet effect \cite{ratchet} and current rectification \cite{rectifier}, which may alter the large on-off current ratios of the PAM-based switch. In order to clarify these effects, we explore laser-driven transport through a two-terminal asymmetric molecular junction with different frequencies. 

\begin{figure}
  \centering
  \includegraphics[width=8cm]{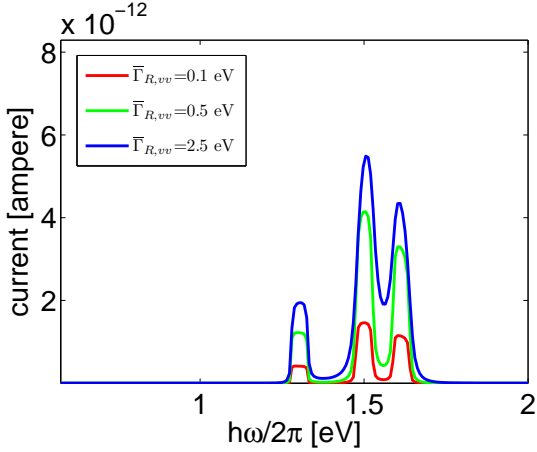}
  \caption{Current-frequency characteristics of a two-terminal PAM junction with asymmetric molecule-lead couplings for $V_{\textrm{SD}}=0.05$ V, $k_B\theta=5\times 10^{-4}$ eV, $\Theta=0$, $E=10^5$ eV, $\overline{\Gamma}_{\textrm{L},uu}=0.5$ eV, $\mu_{\textrm{L}}=\alpha-\textrm{e}V_{\textrm{SD}}/2$, and $\mu_{\textrm{R}}=\alpha+\textrm{e}V_{\textrm{SD}}/2$.}
  \label{fig4}
\end{figure}

Based on the argument in section \ref{section3-1}, the asymmetric effects arising from the electrodes and linker groups can be modeled as molecule-lead coupling functions with different values, $\overline{\Gamma}_{\textrm{L},uu} \neq \overline{\Gamma}_{\textrm{R},vv}$. Fig.~\ref{fig4} shows that the peaks of all lines are located at the same frequencies ($\approx 1.30$, $1.50$, and $1.61$ eV), indicating that the existence of these peaks caused by one-photon assisted tunneling do not change with asymmetric molecule-lead couplings. The asymmetric molecule-lead couplings only affect the peak heights and widths. Note that our analysis is based on  the wide band limit approximation, i.e., we do not consider the effect of the real part of self-energy. However, for a large molecule, such as PAM, the energy shifting caused by the intermediate couplings should not significantly change the electronic properties. As a result, we conclude that the large on-off current ratios of the PAM-based switch should be robust to asymmetry in the molecular junction.

\subsection{Polarization Effect}\label{section3-4}

The field polarization can play a crucial role in the control of electron dynamics. Recently, several theoretical and experimental studies have shown that the aromaticity of benzene \cite{arom1,arom2}, aromatic ring currents in ring-shaped molecules \cite{ring_current_1,ring_current_2,ring_current_3,ring_current_4}, and electron localization in molecular dissociation \cite{localization} can be manipulated by changing the laser polarization. However, these studies focus on the influence of strong linearly or circularly polarized light on isolated molecules, not a molecular junction, which motivates us to explore electron transport through a molecular junction in a weak laser field with linear polarization in various directions.

\begin{figure}
  \centering
  \includegraphics[width=8cm]{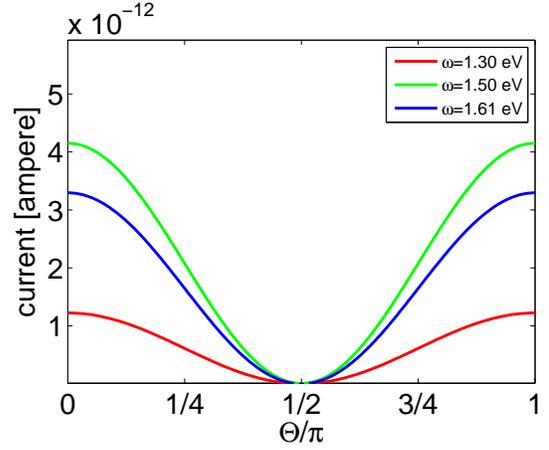}
  \caption{Current as a function of the laser polarization angle $\Theta$ in a two-terminal PAM junction for $V_{\textrm{SD}}=0.05$ V, $k_B\theta=5\times 10^{-4}$ eV, $E=10^5$ V/cm, $\overline{\Gamma}_{\textrm{L},uu}=\overline{\Gamma}_{\textrm{R},vv}=0.5$~eV, $\mu_{\textrm{L}}=\alpha-\textrm{e}V_{\textrm{SD}}/2$, and $\mu_{\textrm{R}}=\alpha+\textrm{e}V_{\textrm{SD}}/2$.}
  \label{fig5}
\end{figure}

Fig.~\ref{fig5} depicts the current through a two-terminal PAM junction in a weak field with different frequencies as a function of laser polarization angle $\Theta$. The light with the frequencies $1.30$, $1.50$, and $1.61$ eV can result in one-photon assisted tunneling. All of the results show the same characteristics: the current reaches the minimum at the laser polarization angle $\Theta=\pi/2$ and the maximum at $\Theta=0$ and $\pi$. The current maximum occurs at $\Theta=0$ and $\pi$ because the oscillation of the electric field can assist electron tunneling when the directions of electron transport and laser polarization are the same, i.e., along the x-direction in Fig.~\ref{fig1}. On the other hand, when the directions of electron transport and laser polarization are vertical, i.e., $\Theta=\pi/2$, the oscillation of the electric field cannot assist electron tunneling, leading to the minimum electric current.

\begin{figure}
  \centering
  \includegraphics[width=8cm]{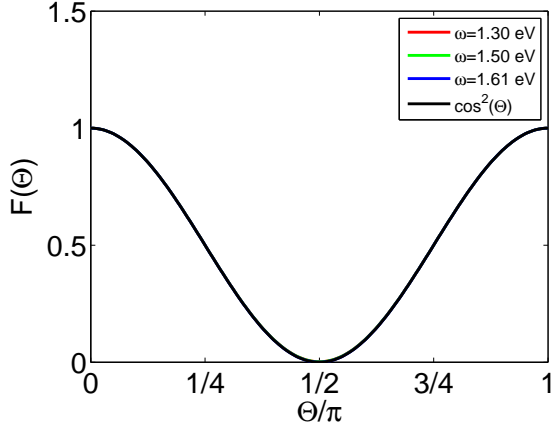}
  \caption{Normalized current as a function of $F(\Theta)$ in a two-terminal PAM junction.}
  \label{fig6}
\end{figure}

Due to all of the curves in Fig.~\ref{fig5} with similar structure, we define the time-averaged current as $\overline{I}_{0}(\omega)$ in a laser field with $\omega$ and at $\Theta=0$, and the factor $F(\omega,\Theta)$ describes the relation between current and laser polarization angle,
\begin{align}
\label{Polarization_angle}
\overline{I}=\overline{I}_{0}(\omega)F(\omega,\Theta).
\end{align}
Fig.~\ref{fig6} shows that all normalized current curves $\overline{I}/\overline{I}_{0}(\omega)$ coincide $\cos^2(\Theta)$, i.e., 
\begin{align}
\label{Polarization_F}
F(\omega,\Theta)=\cos^2(\Theta).
\end{align}
The independence of $F(\omega,\Theta)$ upon frequency is interesting and deserves further exploration. The $\cos^2(\Theta)$ behavior can be understood by the perturbation analysis in Appendix \ref{appendix2}. From Eq.~(\ref{H1_element}), the one-photon Green's function $G^{(\pm 1)}_{vu}(\epsilon)$ is proportional to $\cos(\Theta)$, because in a weak field only the $x$ component of the electric field can assist electron tunneling and the $y$ component cannot. Therefore, the cosine squared relation comes from $\overline{I} \propto |G^{(\pm 1)}_{vu}(\epsilon)|^2 \propto \cos^2(\Theta)$. According to the results shown in Fig.~\ref{fig5}, we can conclude that the on-off switching ratios are sensitive to the laser polarization, indicating that the current magnitude can be conveniently manipulated this way in the laboratory.

\section{Multi-Terminal Network System}\label{section4}

Our methodology is also applicable to multi-terminal systems. To illustrate the methodology, we give two examples, a single-molecule optoelectronic router and a molecular parallel circuit. The former is a three-terminal system connected to three electrodes and the latter is a four-terminal system connected to two electrodes. The molecular networks exhibit a variety of novel and interesting physical phenomena.

\subsection{Single-Molecule Optoelectronic Router}\label{section4-1}

\begin{figure}
  \centering
  \includegraphics[width=8cm]{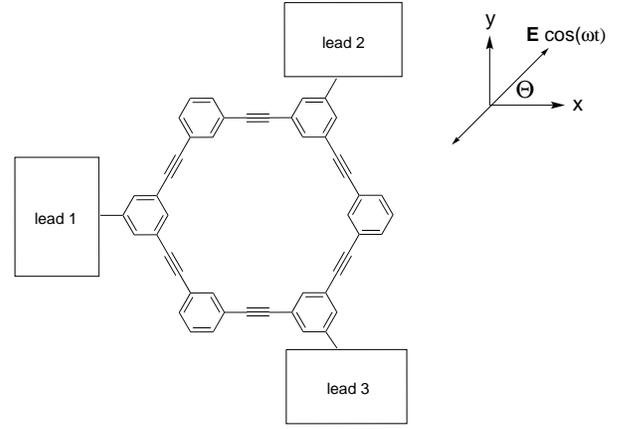}
  \caption{The xy-plane projection of a single-molecule optoelectronic router, which consists of three electrodes (lead 1, lead 2, and lead 3), PAM, and a monochromatic laser field with field strength E, frequency $\omega$, and polarization angle $\Theta$. The chemical potentials are set as $\mu_{\textrm{1}}=\alpha-\textrm{e}V_{\textrm{SD}}$, and $\mu_{\textrm{2}}=\mu_{\textrm{3}}=\alpha$, resulting in lead 2 and 3 having a reflection symmetry along the x axis.}
  \label{fig7}
\end{figure}

Inspired by an analogy from electrical engineering, we propose a new type of current router and explore the correlation between the direction of laser polarization and current. A single-molecular optoelectronic router as shown in Fig.~\ref{fig7} composed of a PAM molecule connected to three electrodes in the presence of a monochromatic laser field. The Hamiltonian is described in Eqs.~(\ref{molHamiltonian}) and (\ref{polarization}), and we assume the coupling functions are $\overline{\Gamma}_{1}=|u\rangle \overline{\Gamma}_{1,uu}\langle u|$, $\overline{\Gamma}_{2}=|v\rangle \overline{\Gamma}_{2,vv}\langle v|$, and $\overline{\Gamma}_{3}=|w\rangle \overline{\Gamma}_{3,ww}\langle w|$, where $|u\rangle$, $|v\rangle$, and $|w\rangle$ denote the $p_z$-orbital on the contact carbon atoms $u$, $v$, and $w$, respectively.

For a three-terminal system, by virtue of Eq.~(\ref{elec_current02}), the current in terminal (lead) 1, 2, and 3 can be expressed as 
\begin{align}
\label{three_current1}
\overline{I}_{1}=\frac{\textrm{e}}{h}\sum_{k=-\infty}^{\infty}\int^{\infty}_{-\infty}d\epsilon [T^{(k)}_{21}(\epsilon)f_{1}(\epsilon)-T^{(k)}_{12}(\epsilon)f_{2}(\epsilon) \nonumber \\
 +T^{(k)}_{31}(\epsilon)f_{1}(\epsilon)-T^{(k)}_{13}(\epsilon)f_{3}(\epsilon)],
\end{align}
\begin{align}
\label{three_current2}
\overline{I}_{2}=\frac{\textrm{e}}{h}\sum_{k=-\infty}^{\infty}\int^{\infty}_{-\infty}d\epsilon [T^{(k)}_{12}(\epsilon)f_{2}(\epsilon)-T^{(k)}_{21}(\epsilon)f_{1}(\epsilon) \nonumber \\
 +T^{(k)}_{32}(\epsilon)f_{2}(\epsilon)-T^{(k)}_{23}(\epsilon)f_{3}(\epsilon)],
\end{align}
\begin{align}
\label{three_current3}
\overline{I}_{3}=\frac{\textrm{e}}{h}\sum_{k=-\infty}^{\infty}\int^{\infty}_{-\infty}d\epsilon [T^{(k)}_{13}(\epsilon)f_{3}(\epsilon)-T^{(k)}_{31}(\epsilon)f_{1}(\epsilon) \nonumber \\
 +T^{(k)}_{23}(\epsilon)f_{3}(\epsilon)-T^{(k)}_{32}(\epsilon)f_{2}(\epsilon)],
\end{align}
where $T^{(k)}_{\alpha \alpha'}(\epsilon)=\textrm{Tr}[\Gamma_{\alpha}(\epsilon+k\hbar\omega)G^{\textrm{R}(k)}(\epsilon)\Gamma_{\alpha'}(\epsilon)G^{\textrm{A}(k)}(\epsilon)]$ for $\alpha$ and $\alpha'=1$, $2$, and $3$. Note that the system does not have generalized parity symmetry, i.e., $T^{(k)}_{\alpha \alpha'}(\epsilon) \neq T^{(k)}_{\alpha' \alpha}(\epsilon)$.

\begin{figure}
  \centering
  \includegraphics[width=8cm]{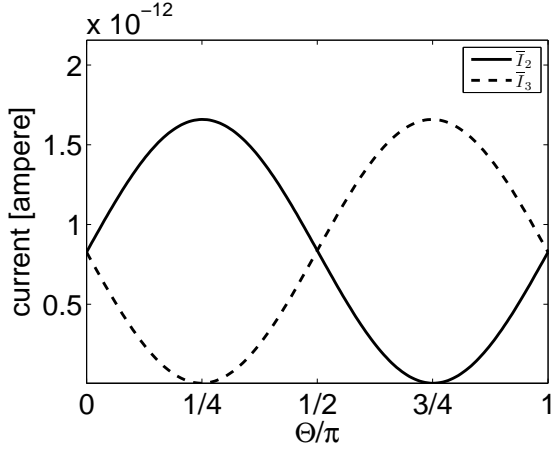}
  \caption{Current as a function of laser polarization angle $\Theta$ in a three-terminal PAM junction for $V_{\textrm{SD}}=0.05$ V, $k_B\theta=5\times 10^{-4}$ eV, $E=10^5$ V/cm, $\hbar\omega=1.30$ eV, and $\overline{\Gamma}_{\textrm{1},uu}=\overline{\Gamma}_{\textrm{2},vv}=\overline{\Gamma}_{\textrm{3},ww}=0.5$~eV.}
  \label{fig8}
\end{figure}

Fig.~\ref{fig8} shows that $\overline{I}_{2}$ and $\overline{I}_{3}$ are functions of laser polarization angle $\Theta$, and $\overline{I}_{2}$ and $\overline{I}_{3}$ are equal at $\Theta=0$, $\pi/2$, and $\pi$. At $\Theta=0$, $\pi/2$, and $\pi$, lead 2 and 3 have reflection symmetry along the x axis in Fig.~\ref{fig7}. However, for any other polarization angle, $\overline{I}_{2}$ and $\overline{I}_{3}$ differ from each other since the reflection symmetry is destroyed by the laser polarization. Moreover, at $\Theta=\pi/4$, $\overline{I}_{2}$ reaches its maximum value while $\overline{I}_{3}$ reaches its minimum value. The maximum value of the current ratio, $\overline{I}_{2}/\overline{I}_{3} \approx 342$, is three times higher than the value reported in a four-site model \cite{optorouter}. In addition, the maximum current ratio $\overline{I}_{3}/\overline{I}_{2}$ at $\Theta=3\pi/4$ is also approximately 342 due to the reflection symmetry of lead 2 and 3. Note that the current ratio depends not only on the symmetry of the router but on laser frequency.

The above results illustrate the correlation between laser polarization and the current ratio, and show that it is possible to manipulate the direction of electric current by using a weak linearly polarized laser field. A single-molecule router is a new type of device, and its utility deserves further exploration.

\subsection{Molecular Parallel Circuits}\label{section4-2}

Series and parallel circuit elements are basic electrical network components. Series circuits at the nanoscale, such as electron transport through linear arrays of quantum dots \cite{ratchet,rectifier,shot_noise_1,shot_noise_2} and alkyl monolayers \cite{Reed_Review,alkyl2,alkyl3}, have been extensively studied, but parallel circuits at the nanoscale have not received such attention. Thus, to bolster the foundation of molecular electronics, an understanding of parallel circuits at the molecular level is important. In this section, we investigate electron transport through a parallel circuit based on a simple two-level model in the presence of a high-frequency driving field.   

\begin{figure}
  \centering
  \includegraphics[width=8cm]{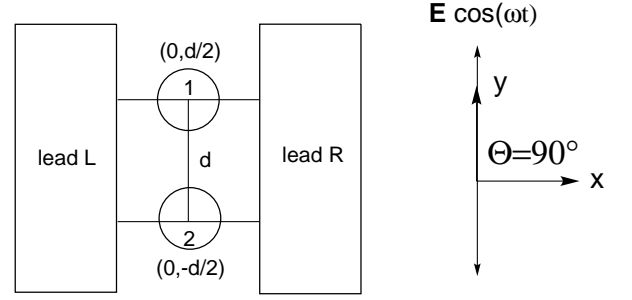}
  \caption{Parallel circuit made of molecular quantum dots 1 and 2 connected to two leads in a laser field with amplitude E, frequency $\omega$, and polarization angle $\Theta=\pi/2$. Here, $d/2$ and $-d/2$ represent the positions of the two dots.}
  \label{fig9}
\end{figure}

Fig.~\ref{fig9} shows a parallel circuit composed of double molecular quantum dots, each of which is connected two leads in the presence of a monochromatic laser field with strength E, frequency $\omega$, and polarization angle $\Theta=\pi/2$, i.e., $\textbf{E}(t)=\textrm{E}\cos(\omega t)\hat{y}$. The positions of the two molecular quantum dots are $\mathbf{r}_{1}=d/2~\hat{y}$ and $\mathbf{r}_{2}=-d/2~\hat{y}$. For simplicity, the molecular Hamiltonian $H_{\textrm{mol}}(t)$ is described by a tight-binding model within the electric dipole approximation,
\begin{align}
\label{Hmatrix3_1}
H_{\textrm{mol}}(t)
&=\left(
\begin{matrix} 
A\cos(\omega t) & \Delta \\
\Delta & -A\cos(\omega t)  
\end{matrix}
\right) \\
\label{Hmatrix3_2}
&=\Delta \sigma_x + A\cos(\omega t)\sigma_z, 
\end{align}
where $A=d\cdot\textrm{E}/2$ corresponds to the electric dipole interaction induced by the laser field, $\Delta$ is the hopping integral between the two quantum dots, and $\sigma_x$, $\sigma_y$, and $\sigma_z$ are Pauli matrices. To facilitate the analysis, we assume that the on-site energy of the quantum dots is zero.

Within the wide band limit approximation and the assumption of symmetric molecule-lead coupling, the coupling function is
\begin{align}
\label{Coupmatrix3_2}
\overline{\Gamma}_{\textrm{L}} = \overline{\Gamma}_{\textrm{R}} =
\left(
\begin{matrix} 
\Gamma & 0\\
0 & \Gamma  
\end{matrix}
\right). 
\end{align}
In order to reduce the complexity of the problem, we do not consider the off-diagonal terms of the coupling function, which may cause cooperative effects \cite{cooperative1,cooperative2,cooperative3}.

By Substitution of Eq.~(\ref{Hmatrix3_1}) and Eq.~(\ref{Coupmatrix3_2}) into Eq.(\ref{transmission}), the retarded Green's function can be divided into four components
\begin{align}
\label{4component1}
T^{(k)}_{\textrm{RL}}(\epsilon)=\Gamma^2(|G^{\textrm{R}(k)}_{11}(\epsilon)|^2+G^{\textrm{R}(k)}_{12}(\epsilon)G^{\textrm{A}(k)}_{21}(\epsilon) \nonumber \\ 
+G^{\textrm{R}(k)}_{21}(\epsilon)G^{\textrm{A}(k)}_{12}(\epsilon)+|G^{\textrm{R}(k)}_{22}(\epsilon)|^2), 
\end{align}
where $T^{(k)}_{\textrm{RL}}(\epsilon)=T^{(k)}_{\textrm{LR}}(\epsilon)$ due to the system with generalized parity symmetry.

\begin{figure}
  \centering
  \includegraphics[width=8cm]{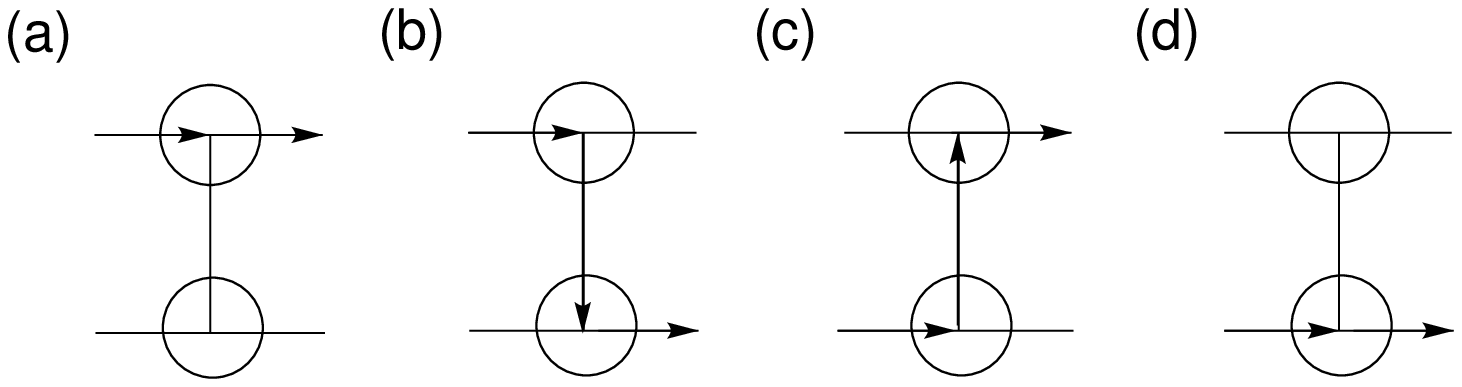}
  \caption{Schematic representation of the four current components. (a) $\overline{I}_{\textrm{R},11}$. (b) $\overline{I}_{\textrm{R},12}$. (c) $\overline{I}_{\textrm{R},21}$. (d) $\overline{I}_{\textrm{R},22}$.}
  \label{fig10}
\end{figure}

Using $G^{\textrm{R}(k)}_{12}(\epsilon)=G^{\textrm{R}(k)}_{21}(\epsilon)$ and substituting Eq.~(\ref{4component1}) into Eq.~(\ref{elec_current02}), the total current can be decomposed into four current components,
\begin{align}
\label{current_serial1}
\overline{I}_{\textrm{R}}=\overline{I}_{\textrm{R},11}+\overline{I}_{\textrm{R},12}+\overline{I}_{\textrm{R},21}+\overline{I}_{\textrm{R},22},
\end{align}
where
\begin{align}
\label{current_serial2}
\overline{I}_{\textrm{R},uv}=\frac{\textrm{e}\Gamma^2}{h}\sum_{k=-\infty}^{\infty}\int^{\infty}_{-\infty}d\epsilon |G^{\textrm{R}(k)}_{uv}(\epsilon)|^2[f_{\textrm{R}}(\epsilon)-f_{\textrm{L}}(\epsilon)]. 
\end{align}
and $u (v)=1$ or $2$. Note that $\overline{I}_{\textrm{R},11}=\overline{I}_{\textrm{R},22}$ and $\overline{I}_{\textrm{R},12}=\overline{I}_{\textrm{R},21}$ due to $G^{\textrm{R}(k)}_{11}(\epsilon)=G^{\textrm{R}(k)}_{22}(\epsilon)$ and $G^{\textrm{R}(k)}_{12}(\epsilon)=G^{\textrm{R}(k)}_{21}(\epsilon)$. The schematic representation of the four current components is shown in Fig.~\ref{fig10}.

\begin{figure}
  \centering
  \includegraphics[width=8cm]{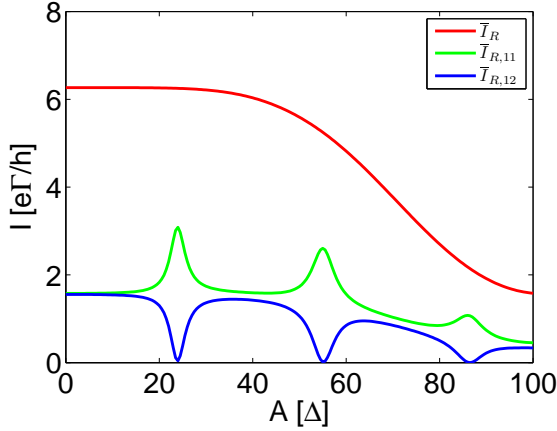}
  \caption{Current-field intensity characteristics of a parallel circuit based on double molecular quantum dots for $V_{\textrm{SD}}= 0.5$ V, $k_{B}T=0$, $\Delta=0.1$ eV, $\hbar\omega=10\Delta$, and $\Gamma=0.1\Delta$. We assume symmetric chemical potentials for the two leads, i.e.,$\mu_{\textrm{L}}=\mu-\textrm{e}V_{\textrm{SD}}/2$ and $\mu_{\textrm{R}}=\mu+\textrm{e}V_{\textrm{SD}}/2$.}
  \label{fig11}
\end{figure}

Fig.~\ref{fig11} shows the current-field intensity characteristics of a parallel circuit based on double molecular quantum dots in a high-frequency driving field ($\hbar\omega=10\Delta$). In the absence of the laser field, i.e., $A=0\Delta$, we can observe $\overline{I}_{\textrm{R},11}\approx\overline{I}_{\textrm{R},12}$ and $\overline{I}_{\textrm{R}}\approx 4\overline{I}_{\textrm{R},11}$, which can be understood quantitatively by the following discussion. First, in the absence of the laser field, Eq.~(\ref{current_serial2}) can reduce to
\begin{align}
\label{current_serial3}
\overline{I}_{\textrm{R},uv}=\frac{\textrm{e}\Gamma^2}{h}\int^{\infty}_{-\infty}d\epsilon |G^{\textrm{R}(0)}_{uv}(\epsilon)|^2[f_{\textrm{R}}(\epsilon)-f_{\textrm{L}}(\epsilon)]. 
\end{align}

For a two-level system, by virtue of Eq.~(\ref{green_zerofield2}), the retarded Green's functions $|G^{\textrm{R}(0)}_{11}(\epsilon)|^2$ and $|G^{\textrm{R}(0)}_{12}(\epsilon)|^2$ have analytical forms and can be approximated as
\begin{align}
\label{serial_green1}
|G^{\textrm{R}(0)}_{11}(\epsilon)|^2=\frac{1}{4}\left| \frac{1}{\epsilon-\Delta-i\Gamma}+\frac{1}{\epsilon+\Delta-i\Gamma} \right|^2 \\
\label{serial_green2}
\approx\frac{1}{4}\left(\frac{1}{(\epsilon-\Delta)^2+\Gamma^2}+\frac{1}{(\epsilon+\Delta)^2+\Gamma^2}\right),
\end{align}
and
\begin{align}
\label{serial_green3}
|G^{\textrm{R}(0)}_{12}(\epsilon)|^2=\frac{1}{4}\left| \frac{1}{\epsilon-\Delta-i\Gamma}-\frac{1}{\epsilon+\Delta-i\Gamma} \right|^2 \\
\label{serial_green4}
\approx\frac{1}{4}\left(\frac{1}{(\epsilon-\Delta)^2+\Gamma^2}+\frac{1}{(\epsilon+\Delta)^2+\Gamma^2}\right).
\end{align}
The cross terms in Eq.~(\ref{serial_green1}) and Eq.~(\ref{serial_green3}) correspond to constructive  and destructive quantum interference between two resonant states, respectively \cite{lineshape}, and they can be neglected in the resonant tunneling regime. Therefore, we have $|G^{\textrm{R}(0)}_{11}(\epsilon)|^2\approx|G^{\textrm{R}(0)}_{12}(\epsilon)|^2$. In addition, for the condition that the energy levels of all resonant states are between $\mu_{\textrm{R}}$ and $\mu_{\textrm{L}}$, we can approximate $\int^{\infty}_{-\infty}d\epsilon |G^{\textrm{R}(0)}_{11}(\epsilon)|^2[f_{\textrm{R}}(\epsilon)-f_{\textrm{L}}(\epsilon)]\approx\int^{\infty}_{-\infty}d\epsilon |G^{\textrm{R}(0)}_{11}(\epsilon)|^2$. We substitute Eqs.~(\ref{serial_green2}) and (\ref{serial_green4}) into Eq.~(\ref{current_serial3}) and make use of $\int^{\infty}_{-\infty}d\epsilon |G^{\textrm{R}(0)}_{11}(\epsilon)|^2=\pi/2\Gamma$, and then we derive $\overline{I}_{\textrm{R},11}\approx\overline{I}_{\textrm{R},12}\approx\pi\textrm{e}\Gamma/2h$ and $\overline{I}_{\textrm{R}}\approx 2\pi\textrm{e}\Gamma/h\approx 4\overline{I}_{\textrm{R},11}$, which are in quantitatively good agreement with the numerical results in Fig.~\ref{fig11}.

In the presence of the laser field ($A\neq 0$), Fig.~\ref{fig11} shows that $\overline{I}_{\textrm{R},12}$ becomes almost zero at $A=24\Delta$, $55\Delta$, and $86\Delta$ while $\overline{I}_{\textrm{R},11}$ reaches local maximum values at the same positions, indicating that the individual current components can be manipulated by a laser field. To understand this behavior, we transform the Hamiltonian in Eq.~(\ref{Hmatrix3_2}) into a rotating frame \cite{TLS1}
\begin{align}
\label{rotating2}
H^{\text{rot}}_{\text{mol}}(t)&=U^{\dagger}_{\textrm{rot}}(t)H_{\text{mol}}(t)U_{\textrm{rot}}(t)-i\hbar U^{\dagger}_{\textrm{rot}}(t)\frac{dU_{\textrm{rot}}(t)}{dt}  \nonumber \\
&=\left(
\begin{matrix} 
0 & \Delta e^{i\frac{A}{\hbar\omega}\sin(\omega t)} \\
\Delta e^{-i\frac{A}{\hbar\omega}\sin(\omega t)}    & 0
\end{matrix}
\right)
\end{align}
by using the operator
\begin{align}
\label{rotating}
U_{\textrm{rot}}(t)=\exp\left(-i\frac{A}{\hbar\omega}\sin(\omega t)\sigma_z\right). 
\end{align}
This transformation gives a good description of the dynamics of a time-periodic system for the high-frequency driving condition $\hbar\omega \gg \Delta$ and the strong-field driving condition $A>\Delta$ hold \cite{TLS1}. Using  that $e^{ia\sin(b)}=\sum^{\infty}_{m=-\infty}J_{m}(a)e^{imb}$ and $J_{-m}(a)=(-1)^mJ_{m}(a)$, where $J_{m}(a)$ are Bessel functions of the first kind and $m$ are integers, we have
\begin{align}
\label{rotating3}
H^{\text{rot}}_{\text{mol}}(t)= \sum\limits_{m=-\infty}^{\infty} \Delta
\left(
\begin{matrix} 
0 &  J_{m}(\frac{A}{\hbar\omega})e^{im\omega t} \\
 (-1)^m J_{m}(\frac{A}{\hbar\omega})e^{im\omega t}    & 0
\end{matrix}
\right).
\end{align} 
Recall that $\hbar\omega\gg\Delta > \Delta|J_{m}(\frac{A}{\hbar\omega})|$ so we can adopt a high-frequency approximation \cite{highf1,highf2} and Eq.~(\ref{rotating3}) can be expressed as a static system with the effective hopping integrals $\Delta_{\textrm{eff}}=\Delta J_{0}(\frac{A}{\hbar\omega})$ as follows
\begin{align}
\label{rotating4}
H^{\text{rot}}_{\text{mol}}&=\left(
\begin{matrix} 
0 & \Delta_{\textrm{eff}} \\
\Delta_{\textrm{eff}}  & 0
\end{matrix}
\right).
\end{align}
Eq.~(\ref{rotating4}) indicates that when $J_{0}(\frac{A}{\hbar\omega})=0$, there is no coupling between two molecular quantum dots, leading to $\Delta_{\textrm{eff}}=0$. Therefore, we can derive the retarded Green's functions in the rotating frame,
\begin{align}
\label{serial_green7}
|\widetilde{G}^{\textrm{R}(0)}_{12(21)}(\epsilon)|^2=\frac{1}{4}\left| \frac{1}{\epsilon-\Delta_{\textrm{eff}}-i\Gamma}-\frac{1}{\epsilon+\Delta_{\textrm{eff}}-i\Gamma} \right|^2 = 0,
\end{align}
leading to $\overline{I}_{\textrm{R},12(21)}=\int^{\infty}_{-\infty}d\epsilon |\widetilde{G}^{\textrm{R}(0)}_{12(21)}(\epsilon)|^2=0$  corresponding to coherent destruction of tunneling \cite{CDT1}. In addition, the first three roots of $J_{0}(\frac{A}{\hbar\omega})$, i.e., $2.405$, $5.520$, and $8.654$, correspond to $A=24.05\Delta$, $55.20\Delta$, and $86.54\Delta$, consistent with the numerical results in Fig.~\ref{fig11}. Moreover, no coupling between the two molecular quantum dots means that the on-site wavefunction amplitude $\tilde{\phi}^{n}_{\nu}=1$ or $0$ in Eq.~(\ref{green_zerofield2}), indicating that current only can pass through the dot 1 or 2. Consequently, $\overline{I}_{\textrm{R},11(22)}$ reaches local maxima at the first three roots of $J_{0}(\frac{A}{\hbar\omega})$. This section showed that the current components of a molecular parallel circuit can be manipulated by a strong high-frequency driving field.


\section{Conclusions and Prospects}

In this study, we develop a new general method to simulate electron transport through a single-molecule junction under time-periodic fields. This method also enables dealing with electron transport through quantum dots and other nanostructures in a strong driving field. To demonstrate the wide range of applications for this methodology, we give three examples: single-molecule optoelectronic switches, routers, and parallel circuits, and investigate their transport properties. Our computations show that PAM-based optoelectronic switches have robust large on-off switch ratios and weak-field operating conditions, which are not sensitive to asymmetric molecule-lead couplings. In addition, the magnitude of the current can be tuned by changing the direction of laser polarization and the field amplitude. The field-amplitude power laws for one- and two-photon assisted tunneling are evident in the computational results, and the laws can be proven by using the perturbation theory. For PAM-based optoelectronic routers, we show that it is possible to manipulate the direction of electric current through the PAM molecule by using a weak linearly polarized laser field. The maximum current ratio $\overline{I}_{2}/\overline{I}_{3}$ depends on the symmetry of the routers, and the ratio can reach approximately 340. For the parallel circuits made of molecular quantum dots, the total current can be divided into four components, the magnitude of which can be controlled by a high-frequency laser field. In addition, we quantitatively derive the value of current in the absence of a laser field and successfully explain the condition of coherent destruction of tunneling by using the rotating wave approximation and the high-frequency approximation. Our study opens up a new direction for exploring light-driven transport through molecular junctions and its potential applications in single-molecule optoelectronics \cite{optoelec,molopto1}.

Although we have developed a new general method to deal with light-driven transport, several issues remain to be resolved. First, our Floquet analysis is based on the wide band limit approximation, i.e., no memory effects. For a practical system, the role of electrodes may be important. Thus, a next step is to develop a more general formulation which enables dealing with the surface Green's function of the electrodes. Second, our method is based on a single-electron model, i.e., an independent electron model, so it does not include many-body effects such as electron-electron interactions and electron-photon interactions \cite{TGU_model}. Lehmann et~al. have investigated vibrational effects in a two-site system \cite{epcoupling}, but they only considered a weak thermal coupling limit, i.e., no many-body effect such as polarons \cite{polaron1,polaron2} in their study. The synergistic effects of many-body interactions and photon-assisted tunneling may exhibit new physical phenomena calling for study. Third, for simplicity we investigate light-driven transport through the three systems using a H\"uckel-type Hamiltonian, but our method can be combined with a more realistic model, e.g., molecular Hamiltonian computed from the density-functional theory in a maximally localized Wannier function representation. High-level simulations may give a fully complete description relevant to experimental investigations of molecular optoelectronics. Fourth, our method cannot deal with electronic excitation in a molecular junction. As the frequencies of light are compatible with the energy difference between the HOMO (highest occupied molecular orbital) and LUMO (lowest unoccupied molecular orbital), photon assisted tunneling and electronic excitation may occur simultaneously. The distinction between the two mechanisms is still an open question. We hope that this work motivates further theoretical and experimental investigations into light-driven transport through single-molecule junctions.

\begin{acknowledgments}
We thank Dr. Tak-San Ho for useful discussions. This research is supported by the NSF (Grant Number
CHE-1058644), ARO (Grant Number W911NF-13-1-0237) and PPST.
\end{acknowledgments}
 
\begin{appendix}


\section{Periodic Charging of the Molecule}\label{appendix1}

Substitution of Eq.~(\ref{rate_green2}) into $\sum_{qn}|\langle n|U(t,t_0)|\alpha q \rangle|^2$ gives 
\begin{align}
\label{charging1}
q_{\alpha}(t)&=\textrm{e} \sum_{qn}|\langle n|U(t,t_0)|\alpha q \rangle|^2f_{\alpha}(\epsilon_{\alpha q}) 
\\
\label{charging2}
&=\frac{\textrm{e}}{2\pi}\int d\epsilon \textrm{Tr}[ G^{\textrm{R}}(t,\epsilon) \Gamma_{\alpha}(\epsilon) G^{\textrm{A}}(t,\epsilon)]f_{\alpha}(\epsilon),
\end{align}
in which $q_{\alpha}(t)$ corresponds to the charge contributed by the lead $\alpha$ to the molecule. It can be proved that $q_{\alpha}(T)=q_{\alpha}(0)$ since $G^{\textrm{R}}(t,\epsilon)$ and $G^{\textrm{A}}(t,\epsilon)$ are periodic functions of time.

Integration of the tunneling rate $k_{n,\alpha q}(t)$ multiplied by the Fermi function $f_{\alpha}(\epsilon_{\alpha q})$ over a period $T$ and use of the fundamental theorem of calculus gives
\begin{align}
\label{charging3}
&\sum_{qn}\int^{T}_{0}dt k_{n,\alpha q}(t)f_{\alpha}(\epsilon_{\alpha q}) \nonumber \\
&=\sum_{qn}\int^{T}_{0}dt\frac{1}{dt}|\langle n|U(t,t_0)|\alpha q\rangle|^2 f_{\alpha}(\epsilon_{\alpha q}) \nonumber \\
&=\sum_{qn} \left(|\langle n|U(T,t_0)|\alpha q\rangle|^2-|\langle n|U(0,t_0)|\alpha q\rangle|^2 \right)f_{\alpha}(\epsilon_{\alpha q}).
\end{align}

By substitution of Eq.~(\ref{charging2}) into Eq.~(\ref{charging3}) and using the relation $G^{\textrm{R(A)}}(t,\epsilon)=G^{\textrm{R(A)}}(t+T,\epsilon)$, we can obtain
\begin{align}
\label{charging4}
\frac{\textrm{e}}{T}\sum_{qn}\int^{T}_{0}dt k_{n,\alpha q}(t) f_{\alpha}(\epsilon_{\alpha q})&= \frac{q_{\alpha}(T)-q_{\alpha}(0)}{T} \nonumber \\
&=0,
\end{align}
indicating that $\textrm{e}\int^{T}_{0}dtk_{n,\alpha q}f_{\alpha}(\epsilon_{\alpha q})$ corresponds to periodic charging of the molecule driven by external time-periodic fields and contributes zero current over a period. 

In addition, we can invoke Eq.~(\ref{rate_green2}) to prove that $k_{\alpha q,n}(t)$ is a periodic function of time. Then, it follows that the first two terms in Eq.~(\ref{current1}) average to zero.

\section{Time-Independent Non-Hermitian Perturbation Theory}\label{appendix2}

Eq.~(\ref{FM}) and its adjoint eigenvalue equation can be expressed as 
\begin{align}
\label{perturbation_H1}
\overline{H}_{F} \phi_{\lambda,\zeta}=q_{\lambda,\zeta}\phi_{\lambda,\zeta}, \\
\label{perturbation_H2}
\overline{H}^{\dagger}_{F} \chi_{\lambda,\zeta}=\overline{q}_{\lambda,\zeta}\chi_{\lambda,\zeta},
\end{align}
where the eigenstates $\phi_{\lambda,\zeta}$ and $\chi_{\lambda,\zeta}$ form a complete biorthogonal basis, i.e., $\chi_{\lambda,\zeta}^{\dagger}\phi_{\lambda,\zeta}=\sum_{nk}\chi^{n,k *}_{\lambda,\zeta}\phi^{n,k}_{\lambda,\zeta}=\sum_{nk}\phi^{n,k}_{\lambda,\zeta}\phi^{n,k}_{\lambda,\zeta}=\delta_{\lambda\lambda'}\delta_{\zeta\zeta'}$, and the eigenvalues $\overline{q}^{*}_{\lambda,\zeta}=q_{\lambda,\zeta}$ \cite{nonHerm1}. In the previous section, the quasienergies $q_{\lambda,0}$ and Floquet states $\phi_{\lambda,0}$ are chosen in the first Brillouin zone, while here for convenience we choose $q_{\lambda,0}$ and $\phi_{\lambda,0}$ which correspond to the energies and the states derived from a Hamiltonian without a driving field, i.e., $q_{\lambda,0}=\tilde{q}_{\nu}$ and $\phi^{n,0}_{\lambda,0}=\tilde{\phi}^{n}_{\nu}$ in Eq.~(\ref{Hamiltonian_zerofield}). 

As the external laser field is weak, we can divide the time-averaged Floquet Hamiltonian into two parts
\begin{align}
\label{perturbation1}
\overline{H}_{F}=\overline{H}_{0}+\Lambda \overline{H}_{1},
\end{align}
where $\Lambda$ is a dimensionless parameter ranging continuously from 0 (zero perturbation) to 1 (the full perturbation), $\overline{H}_{0}$ is a non-Hermitian unperturbed Hamiltonian, and $\overline{H}_{1}$ is a perturbed Hamiltonian. According to Eq.~(\ref{supermatrix}), we separate $\overline{H}_{F}$ into
\begin{align}
\label{H0} 
&\overline{H}_{0}= \nonumber\\
&\left(\begin{matrix}
  \ddots  &   \vdots                       &    \vdots                  &  \vdots   &  \vdots                    &\vdots&  \\   
  \cdots  &   A-2\hbar\omega\textbf{I}     &   0                        &   0       &  0                         &  0   & \cdots    \\
  \cdots  &   0                            &   A-\hbar\omega\textbf{I}  &   0       &  0                         &  0   & \cdots     \\
  \cdots  &   0                            &   0                        &   A       &  0                         &  0   & \cdots     \\
  \cdots  &   0                            &   0                        &   0       &  A+\hbar\omega\textbf{I}   &  0   & \cdots     \\
  \cdots  &   0                            &   0                        &   0       &  0                         &  A+2\hbar\omega\textbf{I}   &\cdots \\ 
         	&   \vdots                       &    \vdots                  &  \vdots   &  \vdots                    &\vdots&   \ddots\\ 
\end{matrix}\right),
\end{align}
and
\begin{align}
\label{H1} 
\overline{H}_{1}= 
\left(\begin{matrix}
  \ddots  &   \vdots     &    \vdots     &  \vdots   &  \vdots                    &\vdots&  \\   
  \cdots  &   0           &   B           &   0       &  0                         &  0   & \cdots    \\
  \cdots  &   B          &   0           &   B       &  0                         &  0   & \cdots     \\
  \cdots  &   0          &   B           &   0       &  B                         &  0   & \cdots     \\
  \cdots  &   0          &   0           &   B       &  0                         &  B   & \cdots     \\
  \cdots  &   0          &   0           &   0       &  B                         &  0  &\cdots&   \\
					&    \vdots    &  \vdots       &  \vdots   &\vdots                      & \vdots & \ddots \\ 
\end{matrix}\right).
\end{align} 

We begin with the unperturbed non-Hermitian Hamiltonian $\overline{H}_{0}$, which has known eigenvalues and eigenstates
\begin{align}
\label{noperturbation_H1}
\overline{H}_{0} \phi^{0}_{\lambda,\zeta}=q^{(0)}_{\lambda,\zeta}\phi_{\lambda,\zeta}^{(0)}, \\
\label{noperturbation_H2}
\overline{H}^{\dagger}_{0} \chi^{(0)}_{\lambda,\zeta}=\overline{q}^{(0)}_{\lambda,\zeta}\chi^{(0)}_{\lambda,\zeta},
\end{align}
which satisfy $\sum_{nk}\chi^{(0) nk *}_{\lambda,\zeta}\phi^{(0)nk}_{\lambda,\zeta}=\delta_{\lambda\lambda'}\delta_{\zeta\zeta'}$ and $\overline{q}^{(0)*}_{\lambda,\zeta}=q^{(0)}_{\lambda,\zeta}$. Note that $\overline{H}_0$ and $\overline{H}^{\dagger}_{0}$ are block diagonal matrices so 
\begin{align}
\label{blockcoef1}
\phi^{(0)n,k}_{\lambda,\zeta}=\phi^{(0)n,k}_{\lambda,\zeta}\delta_{k,\zeta}. 
\end{align}
\begin{align}
\label{blockcoef2}
\chi^{(0)n,k}_{\lambda,\zeta}=\chi^{(0)n,k}_{\lambda,\zeta}\delta_{k,\zeta}.
\end{align}

If $\overline{H}_{1}$ is sufficiently weak, $\phi_{\lambda,\zeta}$ and $q_{\lambda,\zeta}$ can be written as a power series in $\Lambda$,
\begin{align}
\label{perturbation2_1}
\phi_{\lambda,\zeta}&=\phi^{(0)}_{\lambda,\zeta}+\Lambda\phi^{(1)}_{\lambda,\zeta}+\Lambda^2\phi^{(2)}_{\lambda,\zeta}+..., \\
\label{perturbation2_2}
q_{\lambda,\zeta}&=q_{\lambda,\zeta}^{(0)}+ \Lambda q_{\lambda,\zeta}^{(1)}+\Lambda^2 q_{\lambda,\zeta}^{(2)}+..., 
\end{align}

Substituting Eqs.~(\ref{perturbation1}), (\ref{perturbation2_1}), and (\ref{perturbation2_2}) into Eq.~(\ref{perturbation_H1}), expanding Eq.~(\ref{perturbation_H1}) in powers of $\Lambda$, and multiplying by $\chi^{(0) \dagger}_{\lambda',\zeta'}$ give the first-order correction to the coefficients of the wave function 
\begin{align}
\label{perturbation_coeff1}
&\phi^{(1)n,-k}_{\lambda,\zeta}= \nonumber \\
&\sum_{\lambda'\zeta'}\sum_{n'k'}\sum_{n''k''}\frac{\chi^{(0)n'',k'' *}_{\lambda',\zeta'}[\overline{H}_1]_{n''k'',n'k'}\phi^{(0)n',k'}_{\lambda,\zeta}}{q^{(0)}_{\lambda,\zeta}-q^{(0)}_{\lambda',\zeta'}}\phi^{(0)n,-k}_{\lambda',\zeta'},
\end{align}
where $(\lambda',\zeta')\neq(\lambda,\zeta)$. Eq.~(\ref{perturbation_coeff1}) has a similar form to the first-order correction for the coefficients of the wave function derived from standard time-independent perturbation theory. According to Eq.~(\ref{molHamiltonian}), we obtain
\begin{align}
\label{H1_element}
&[\overline{H}_1]_{n''k'',n'k'} \nonumber \\
&=-\frac{\textrm{e}\textrm{E}}{2}(x_{n'}\cos\Theta+y_{n'}\sin\Theta)\delta_{n'',n'}\delta_{k'',k'\pm 1}.
\end{align}

By substitution of $\phi_{\lambda,\zeta}=\phi^{(0) n,k}_{\lambda,\zeta}+\phi^{(1) n,k}_{\lambda,\zeta}$ into Eq.~(\ref{green2}) and by virtue of Eqs.~(\ref{blockcoef1}), (\ref{blockcoef2}), (\ref{perturbation_coeff1}) and (\ref{H1_element}), it is straightforward to derive
\begin{widetext}
\begin{align}
\label{green_pert}
G^{\textrm{R}(\pm 1)}_{nn'}(\epsilon)&=\sum_{\lambda}\sum^{+\infty}_{\zeta=-\infty}\frac{ \phi^{n,\mp 1}_{\lambda,\zeta} \phi_{\lambda,\zeta}^{n', 0}}{\epsilon-q_{\lambda,\zeta}} \\
&\approx \sum_{\lambda}\frac{ \phi^{(0) n,\mp 1}_{\lambda,0} \phi_{\lambda,0}^{(0) n',0}}{\epsilon-q_{\lambda,0}}+
\sum_{\lambda}\frac{ \phi^{(0) n,\mp 1}_{\lambda,\mp 1} \phi_{\lambda,\mp 1}^{(1) n',0}}{\epsilon-q_{\lambda,\mp 1}} +
\sum_{\lambda}\frac{ \phi^{(1) n,\mp 1}_{\lambda,0} \phi_{\lambda,0}^{(0) n', 0}}{\epsilon-q_{\lambda,0}}+
\sum_{\lambda}\sum^{+\infty}_{\zeta=-\infty}\frac{ \phi^{(1) n,\mp 1}_{\lambda,\zeta} \phi_{\lambda,\zeta}^{(1) n', 0}}{\epsilon-q_{\lambda,\zeta}},
\end{align}
\end{widetext}
where the first term is equal to zero and we neglect the fourth term which corresponds to the next-order correction. As a result, we obtain 
\begin{align}
\label{green_pert1}
G^{\textrm{R}(\pm 1)}_{nn'}(\epsilon)&\approx\sum_{\lambda}\frac{ \phi^{(0) n,\mp 1}_{\lambda,\mp 1} \phi_{\lambda,\mp 1}^{(1) n',0}}{\epsilon-q_{\lambda,\mp 1}} +\sum_{\lambda}\frac{ \phi^{(1) n,\mp 1}_{\lambda,0} \phi_{\lambda,0}^{(0) n', 0}}{\epsilon-q_{\lambda,0}},
\end{align}
Suppose the energy of a tunneling electron is close to the energy of a particular one-photon quasistate $S$, i.e., $\epsilon\approx\epsilon_{S}\mp\hbar\omega$, and is far away from the other states. Then, Eq.~(\ref{green_pert1}) becomes
\begin{align}
\label{green_pert1_1}
G^{\textrm{R}(\pm 1)}_{nn'}(\epsilon)&\approx\frac{ \phi^{(0) n,\mp 1}_{S,\mp 1} \phi_{S,\mp 1}^{(1) n',0}}{\epsilon-q_{{S,\mp 1}}},
\end{align}
According to Eqs. (\ref{perturbation_coeff1}) and (\ref{H1_element}), we can show
\begin{align}
\label{green_pert1_2}
G^{\textrm{R}(\pm 1)}_{nn'}(\epsilon)& \propto \phi_{S,\mp 1}^{(1) n',0} \propto \textrm{E},
\end{align}
which indicates that the current induced by one-photon assisted tunneling is proportional to $\textrm{E}^2$. Moreover, by using the second-order correction to the coefficients of the wave function (not shown here), we can derive that 
\begin{align}
\label{green_pert2}
G^{\textrm{R}(\pm 2)}_{nn'}(\epsilon) \propto \textrm{E}^2,
\end{align}
and the current induced by two-photon assisted tunneling is proportional to $\textrm{E}^4$.

Note that Eq.~(\ref{perturbation_coeff1}) is generally valid in the case of $q^{(0)}_{\lambda,\zeta}\neq q^{(0)}_{\lambda',\zeta'}$, e.g., the high frequency limit. However, according to Eqs.~(\ref{blockcoef1}), (\ref{blockcoef2}), and (\ref{H1_element}), the conditions $q^{(0)}_{\lambda,\zeta}\neq q^{(0)}_{\lambda',\zeta'}$ can reduce to $q^{(0)}_{\lambda,\pm 1}\neq q^{(0)}_{\lambda',0}$. We found that all states satisfy $q^{(0)}_{\lambda,\pm 1} \neq q^{(0)}_{\lambda',0}$, so it is reasonable to apply the analysis to establish the field-amplitude power laws.

\end{appendix}

\end{document}